\documentclass[12pt,a4paper]{article}
\pdfoutput=1
\usepackage{color}
\usepackage{amssymb,amsmath,bm,bbm}
\usepackage{epsf}
\usepackage{epsfig}
\usepackage{afterpage}
\usepackage{longtable}
\usepackage[dvipsnames]{xcolor}
\usepackage[linktoc=page,bookmarks=false,colorlinks=false,
linkbordercolor=RoyalBlue,citebordercolor=ForestGreen,urlbordercolor=CornflowerBlue]{hyperref}
\usepackage{latexsym,mathrsfs,dsfont}
\usepackage[normalem]{ulem} % for strikeout \sout
\usepackage[compress]{cite}
\usepackage{graphicx}
\usepackage{url}
\usepackage{paralist}
\usepackage{bbold}

\newcommand{\ord}{\mathcal{O}}

\newcommand{\IM}{{\rm Im}}
\newcommand{\RE}{{\rm Re}}

\newcommand{\gev}{\, {\rm GeV}}

\newcommand{\bsi}{B_6^{(1/2)}}
\newcommand{\bei}{B_8^{(3/2)}}

\def\epe{\varepsilon'/\varepsilon}
\newcommand{\beq}{\begin{equation}}
\newcommand{\eeq}{\end{equation}}
\newcommand{\be}{\begin{equation}}
\newcommand{\ee}{\end{equation}}
\newcommand{\bi}{\begin{itemize}}
\newcommand{\ei}{\end{itemize}}
\newcommand{\ba}{\begin{array}}
\newcommand{\ea}{\end{array}}
\newcommand{\beqa}{\begin{eqnarray}}
\newcommand{\eeqa}{\end{eqnarray}}
\newcommand{\bea}{\begin{eqnarray}}
\newcommand{\eea}{\end{eqnarray}}
\newcommand{\beqn}{\begin{eqnarray}}
\newcommand{\eeqn}{\end{eqnarray}}

\newcommand{\im}{{\rm Im}}

\definecolor{red}{cmyk}{0,1,1,0.4}

\def\kpn{K^+\rightarrow\pi^+\nu\bar\nu}
\def\klpn{K_{L}\rightarrow\pi^0\nu\bar\nu}

\newcommand{\kepe}{\kappa_{\varepsilon^\prime}}

\usepackage{fancyhdr}
\advance \headheight by 3.0truept       % for 12pt mandatory...
\begin{document}

%\begin{flushleft}
%{\em Version of \today}
%{\em Version of 14 September 2015}
%\end{flushleft}

\vspace{-14mm}
\begin{flushright}
        {AJB-21-1}
\end{flushright}

\vspace{8mm}

\begin{center}
{\LARGE\bf
\boldmath{The $\epe$-Story: $1976 - 2021$}
\\[8mm]
{\large\bf Andrzej~J.~Buras \\[0.3cm]}}
{\small 
      TUM Institute for Advanced Study, Lichtenbergstr.~2a, D-85748 Garching, Germany\\
Physik Department, TU M\"unchen, James-Franck-Stra{\ss}e, D-85748 Garching, Germany\\  E-mail: aburas@ph.tum.de }
\end{center}

\vspace{4mm}

\begin{abstract}
  \noindent
The ratio
$\epe$  measures the size of the direct CP violation in $K_L\to\pi\pi$ decays
$(\varepsilon^\prime)$ relative to the indirect one described by $\varepsilon$ and 
is very sensitive to new sources of CP violation. As such it played a prominent 
role in particle physics already for 45 years. Due to the smallness of
 $\epe$ its measurement required  heroic efforts  in the
1980s and the 1990s on both sides of the Atlantic  with final results
presented by  NA48 and KTeV collaborations at the beginning of this millennium.
On the other hand, even 45 years after the first calculation of $\epe$ we
do not know to which degree the Standard Model  agrees with this data and
how
large is the room left for new physics (NP) contributions to this ratio.
This is due to significant non-perturbative (hadronic) uncertainties accompanied by  partial cancellation between the QCD penguin  contributions and electroweak penguin 
contributions. In addition to the calculation of hadronic matrix elements
of the relevant operators including isospin breaking effects and QED corrections, it is crucial to evaluate accurately the Wilson coefficients
of the relevant operators. While the significant control over the latter short distance effects has been achieved already in the early 1990s, with several improvements since then, different views on the non-perturbative contributions to $\epe$
have been expressed by different authors over last thirty years. In fact
even  at the dawn of the 2020s the uncertainty in the room left
for NP contributions to $\epe$ is still very significant, which I find to be very exciting.

My own work on $\epe$ started in 1983 and involved both
perturbative and non-perturbative calculations. This writing is a non-technical recollection of the steps which led to the  present status of $\epe$
including several historical remarks not known to everybody. 
The present status of the $\Delta I=1/2$ rule is also summarized.

This story is dedicated to my great $\epe$-collaborator Jean-Marc G{\'e}rard
on the occasion of the 35th anniversary of our collaboration and his 64th birthday.
\end{abstract}

\setcounter{page}{0}
\thispagestyle{empty}
\newpage

%\tableofcontents

%\newpage

\section{Overture}
Let me open this writing with the 2020 formula  
for $\epe$ within the SM presented in \cite{Aebischer:2020jto}. It reads
\begin{equation}
\left(\frac{\varepsilon'}{\varepsilon}\right)_{\text{SM}} =  
\IM\lambda_{\rm t}\cdot \left[\,
\big(1-\hat\Omega_{\rm eff}\big) \big(-2.9 + 15.4\,\bsi(\mu^*)\big) + 2.0 -8.0\,\bei(\mu^*) \,\right].
\label{AN2020}
\end{equation}
It includes NLO QCD corrections to the QCD penguin  (QCDP) contributions and NNLO contributions to electroweak penguins (EWP). The coefficients in this formula
and the parameters $\bsi$ and $\bei$ are scale dependent. Their values for
different scales are collected in Table~1 of \cite{Aebischer:2020jto}. Here
we will set  $\mu^*=1\gev$ because at this scale it is most convenient to compare
the values for $\bsi$ and $\bei$ obtained in various non-perturbative approaches.
The four contributions in (\ref{AN2020}) are dominated
by the  following operators:
\begin{itemize}
\item
  The terms involving the non-perturbative parameters $\bsi$  and $\bei$ 
contain only the contributions from the dominant QCDP  operator
$Q_6$ and the dominant EWP operator $Q_8$, respectively. There are two main reasons
why $Q_8$ can compete with $Q_6$ here despite the smallness of the electroweak couplings relative to the QCD one. In the basic formula for $\epe$ its contribution is 
enhanced relative to the $Q_6$'s one by the  factor ${\RE A_0/\RE A_2}=22.4$ with  $A_{0,2}$ being isospin amplitudes. In addition its Wilson coefficient is enhanced for the large top-quark mass which is not the case of the $Q_6$'s one. The expressions for
these two operators and the remaining operators mentioned below are given in Appendix~\ref{OPE}. 
\item
The term $-2.9$ is fully dominated by the QCDP operator $Q_4$.
\item
  The term $+2.0$ is fully dominated by EWP operators $Q_9$ and $Q_{10}$.
\item
  The quantity $\hat\Omega_{\rm eff}$ represents the isospin breaking corrections
  and QED corrections beyond EWP contributions. It is not in the ballpark of a few percent  as one
  would naively expect,  because in $\epe$ it is 
enhanced by the  factor ${\RE A_0/\RE A_2}=22.4$.
\item
${\IM\lambda_{\rm t}}$ is the CKM factor that within a few percent is in the ballpark of ${1.45\cdot 10^{-4}}$.
\end{itemize}

On the other hand 
 the experimental world average of $\epe$ 
from NA48 \cite{Batley:2002gn} and KTeV
\cite{AlaviHarati:2002ye,Abouzaid:2010ny} collaborations reads
\be\label{EXP}
(\epe)_\text{exp}=(16.6\pm 2.3)\times 10^{-4} \,.
\ee

In what follows we will describe the theoretical developments
which led over four decades to the
present values of $\bsi$, $\bei$ and  $\hat\Omega_{\rm eff}$ obtained
in various non-perturbative approaches. Inserting them
into (\ref{AN2020}) will allow us to get various expectations for $\epe$
in the SM and to compare them with the data in (\ref{EXP}).

\boldmath
\section{The First Period: $1976-1989$}\label{P1}
\unboldmath
The story of $\epe$ begins on February~1  1976 when John Ellis, Mary K. Gaillard and Dimitri
Nanopoulos submitted a paper to Nuclear Physics B in which the first
calculation of $\epe$ has been presented
\cite{Ellis:1976fn}. This pioneering calculation does not resemble by any means the
present calculations of $\epe$ but this was the first one. In particular, it does not
include renormalization group effects and takes only QCDP
contributions into account. Moreover in 1976 one had no idea about the matrix elements
of QCDP operators contributing to $\epe$.

However, already in 1975, Shifman, Vainshtein and Zakharov \cite{Shifman:1975tn}, suggested that the QCDPs could be responsible for the $\Delta I=1/2$ rule: a large
enhancement of the $K\to\pi\pi$ isospin amplitude $A_0$ over $A_2$ one as seen already above. This was quite
natural at that time because QCDPs contribute only to $A_0$ and the
attempts in 1974 to explain this rule through current-current operators ($Q_1$ and $Q_2$)
including leading order QCD corrections in \cite{Altarelli:1974exa,Gaillard:1974nj} turned out to be unsuccessful. As found twelve years later 
in the framework of the Dual QCD (DQCD) approach \cite{Bardeen:1986vz} and recently confirmed by the RBC-UKQCD lattice QCD collaboration \cite{Abbott:2020hxn},  this failure was caused by 
the poor knowledge of hadronic matrix elements of these operators
at that time. We will be more explicit about it at the end of this writing.

In order to explain the  $\Delta I=1/2$ rule with the help of QCDPs
Shifman, Vainshtein and Zakharov have simply chosen the values of their
hadronic matrix elements so that this rule could be reproduced.
But the same matrix elements enter $\epe$ and Gilman and Wise 
\cite{Gilman:1978wm} using their values from the Russian trio and performing renormalization
group analysis for $m_t$ much smaller than $M_W$, as thought in 1979,
found $\epe$ to be in the ballpark of $5\times 10^{-2}$. Soon after other analyses appeared, in particular the one by Guberina and Peccei in \cite{Guberina:1979ix}, finding values of $\epe$  in the ballpark of $10^{-3}-10^{-2}$.

My first real encounter with $\epe$ goes back to 1981 when Bruce Winstein, the spokesman of future E731 and KTeV collaborations, entered my office at Fermilab and asked me about the work of Gilman and Wise. I told
him that I never studied this ratio, but that one should seriously consider
their result after they wrote other important papers, in particular the one in \cite{Gilman:1979bc}. Bruce told me that if $\epe$ was as large as claimed
by Fred and Mark, he can definitely measure it and I encouraged him to do
it because of the importance of CP violation.

My first paper on $\epe$, in collaboration with Slominski and Steger, appeared
in 1983 \cite{Buras:1983ap}. It was just a phenomenology of this ratio
as a function of the hadronic matrix element of $Q_6$, of the CKM parameters
and of  $m_t$ in the ballpark of $30\gev$ as expected at that time.

Over the 1980s the calculations of $\epe$
were refined through the inclusion of isospin breaking in the
quark masses \cite{Donoghue:1986nm,Buras:1987wc},
the inclusion of QED penguin effects for $m_t \le M_W$
\cite{Bijnens:1983ye,Donoghue:1986nm,Buras:1987wc,Sharpe:1987cx,Lusignoli:1988fz}
and in particular through the first calculation of hadronic matrix elements
of QCDP and EWP operators in QCD by Bardeen, G{\'e}rard and myself in
the framework of the DQCD \cite{Bardeen:1986vp,Bardeen:1986uz,Bardeen:1986vz}.
The latter calculations were done in the strict large $N$ limit of colours, but this
was sufficient to see that QCDPs are not responsible for the $\Delta I=1/2$ rule
\footnote{More about it in Section~\ref{Sec6}.}
and that $\epe$ was rather $\ord(10^{-3})$ than $\ord(10^{-2})$ as claimed
by Gilman and Wise in \cite{Gilman:1978wm}. Thus already in 1986 the values
of $\bsi$ and $\bei$ were known:
\be\label{LN}
\bsi=1.0, \qquad \bei=1.0, \qquad (N \to \infty)
\ee
and the study of their renormalization scale dependence for $\mu \ge 1\gev$ in
the leading logarithmic (LO) approximation 
indicated that it was small, although as we will see later it is non-negligible
when higher order QCD corrections are taken into account and scales below
$1\gev$ in the framework of DQCD are considered.

It is rather surprizing that we did not calculate these two parameters
including $1/N$ corrections in the 1980s.  Such a calculation has been performed by
Jean-Marc G{\'e}rard and me 29 years later with interesting consequences. I  will report on it in Section~\ref{P4}.

While the numerical coefficients  in (\ref{AN2020}) include higher order
QCD corrections that were unknown in the 1980s, it is tempting
to use this formula for $\bsi=1.0$ and $\hat\Omega_{\rm eff}=0.29$ setting
the last two terms to zero because EWP are irrelevant for low values of
$m_t$ expected at that time. Setting ${\IM\lambda_{\rm t}}=1.45\cdot 10^{-4}$
we find
\be
\left(\frac{\varepsilon'}{\varepsilon}\right)_{\text{SM}}= 12.9\times 10^{-4}\,,
\ee
in a good agreement with (\ref{EXP}) which was first known fifteen years
later. Setting $\hat\Omega_{\rm eff}=0$ would increase this value to $18.1\times 10^{-4}$.
This is also consistent with experiment but this change illustrates
the importance of isospin breaking corrections in the evaluation of $\epe$.
We will return to it in Section~\ref{P4}, where we will explain the dynamics
behind $\hat\Omega_{\rm eff}=0.29$. There is no point in discussing the
errors in these estimates already now. We will do it in Section~\ref{P5}.

One of the last analyses of $\epe$ in the SM for $m_t\le M_W$, that
included all effects known at that time, is the one in \cite{Buras:1987qa}
in which typical values for $\epe$ were found in the ballpark of a few $10^{-3}$.

In December 1988 I attended the Kaon conference in Vancouver. Fred Gilman during
a lunch told me about his paper with Claudio Dib and Isard Dunietz \cite{Dib:1988md}, in  which they calculated electroweak contributions to $K_L\to\pi^0e^+ e^-$ for an arbitrary top quark mass finding for $m_t> 150\gev$
a large contribution from the $Z^0$-penguin that increased with $m_t$ roughly
 as $m_t^2$. A similar result has been obtained by Jonathan Flynn and Lisa Randall \cite{Flynn:1988ve}.  This gave me the idea to calculate  $Z^0$-penguin contribution  to $\epe$ for large $m_t$. The QCDP and photon penguin
contributions were known already at that time to have a very weak $m_t$ dependence.

I moved to TUM in November 1988, but already in January 1989 I got a number
of very good diploma students. Two of them, Gerhard Buchalla and Michala
Harlander were supposed to perform the calculation of $\epe$ together with me
for an arbitrary top quark mass. But this was their first calculation of that
type and I was simultaneously busy preparing first lectures in my life
 as well  starting the project on NLO QCD corrections to all flavour
violating processes in collaboration with Peter Weisz \cite{Buras:1989xd}. 
Consequently my project with Gerhard and Michaela took longer than
I initially expected and in April 1989\footnote{The paper appeared already in March but in 1989 the arXiv did not exist and we learned about it several weeks later.} we were surprized by a paper by
Flynn and Randall  \cite{Flynn:1989iu} in which $\epe$ including 
 QCDP,  EWP ($\gamma$ and $Z^0$ penguins)
and the relevant box diagrams were calculated for an arbitrary
top quark mass. Significant suppression of
$\epe$ for $m_t > 150\gev$ by EWPs has been found by them. While definitely Flynn and Randall
should be given the credit for pointing out the importance of $Z^0$ penguins
in $\epe$ in print, fortunately for us their calculation of the QCDP contribution for large $m_t$
was incorrect so that in fact the first correct calculation of $\epe$ 
in the SM for arbitrary $m_t$ including  LO QCD corrections has
been presented by us in \cite{Buchalla:1989we}. In fact the suppression of
$\epe$ for large $m_t$ turned out to be significantly stronger
than reported by Jonathan and Lisa in \cite{Flynn:1989iu}.
The strong cancellation between QCDP
and EWP for $m_t > 150~\gev$ found by us was soon confirmed in the erratum
to \cite{Flynn:1989iu} and subsequently by other authors \cite{Paschos:1991as,Lusignoli:1991bm}. In fact at that 
time, due to the aforementioned cancellation between QCDP and EWP contributions, the vanishing of $\epe$ in the SM and negative values for it could not be excluded.

It is then tempting to include next the EWP terms in (\ref{AN2020}) valid for the presently known $m_t$
and set $\bsi=\bei=1.0$ and $\hat\Omega_{\rm eff}=0.29$ to find this time
\be\label{42}
\left(\frac{\varepsilon'}{\varepsilon}\right)_{\text{SM}}= 4.2\times 10^{-4},
\ee
that is by a factor of $4$   below the experimental value in (\ref{EXP}).
Setting $\hat\Omega_{\rm eff}=0$ would increase this value to $9.4\times 10^{-4}$,
roughly by a factor of two below the data and showing the importance
of isospin breaking effects dominating the parameter  $\hat\Omega_{\rm eff}$. 

As the upper bound on $m_t$ from electroweak precision tests was still
unknown in 1989, one can find  in our paper plots of $\epe$ as a function
of $m_t$ with $\epe$ vanishing for $m_t\approx 200\gev$. I was visiting
CERN in November 1989, one month after our paper appeared. In the CERN
cafeteria  I met Jack Steinberger and asked him the standard question
how he was doing. He told me that NA31 collaboration was shocked by
the theory. I asked him which theory. Your paper with Buchalla and
Harlander he replied. In fact at that time only the first result from NA31
collaboration was known \cite{Burkhardt:1988yh} implying $\epe=(33\pm 11)\times 10^{-4}$. The second NA31 analysis in 1992 gave  $\epe=(23\pm 7)\times 10^{-4}$
\cite{Barr:1993rx} and
the one from the E731 experiment at Fermilab  $\epe=(7.4\pm 5.9)\times 10^{-4}$ \cite{Gibbons:1993zq}. Therefore, the experimental situation of $\epe$ was still unclear in the first half
of the 1990s.

It should be emphasized that the calculations described above 
were done in the leading logarithmic
approximation (e.g.\ one-loop anomalous dimensions of the relevant
operators) with the exception of the $m_t$-dependence which in 
the analyses  \cite{Flynn:1989iu,Buchalla:1989we,Paschos:1991as} has been already
included at the NLO level. While such a procedure is not fully
consistent, $m_t$ dependence enters $\epe$ first at NLO in the renormalization group (RG) improved 
perturbation theory, it allowed for the first time to exhibit the strong
$m_t$-dependence of the EWP contributions and to identify
the partial cancellation between QCDP and EWP contributions 
which is not seen in the strict leading logarithmic approximation.

\boldmath
\section{The Second Period: $1990-2000$}\label{P2}
\unboldmath
During the early 1990s considerable progress has been made by
calculating complete NLO corrections to $\epe$. This means both QCD and electroweak corrections at NLO in the RG improved perturbation theory. This rather heroic effort has been accomplished  by the  Munich and Rome teams led by me \cite{Buras:1991jm,Buras:1992tc,Buras:1992zv,Buras:1993dy} and Guido Martinelli \cite{Ciuchini:1992tj,Ciuchini:1993vr}, respectively. I stress
{\em heroic} because in that days the technology for two-loop calculations
in weak decays, involving issues of $\gamma_5$ and evanescent operators, was
at an early stage and moreover some of my colleagues told me that I was crazy
doing such calculations in view of poorly known hadronic matrix elements. But while the main goal in these papers was to calculate Wilson coefficients of
QCDP and EWP operators at the NLO level in the context of $\epe$, as a
byproduct, many  of our results could also be soon used for non-leptonic $B$ and $D$ decays.
Reviews of this period of
NLO calculations can be found in \cite{Buchalla:1995vs,Buras:2011we} and in my recent book on weak decays \cite{Buras:2020xsm}.

Thus by 1993 $\epe$ was known at the NLO level as far as the Wilson coefficients of the contributing operators are concerned. 
On the other hand, the matrix
elements of the contributing operators were either calculated in the large
$N$ limit, represented for $Q_6$ and $Q_8$ operators by (\ref{LN}), or vacuum insertion method which for these two operators gives basically the same result as  the large $N$ limit\footnote{It should be stressed that the vacuum insertion method has no QCD basis and gives even wrong signs for the $1/N$ corrections to hadronic matrix elements as pointed out already in \cite{Buras:1985yx,Bardeen:1986vp,Bardeen:1986uz,Bardeen:1986vz} and confirmed numerically by the RBC-UKQCD collaboration in \cite{Boyle:2012ys}.}.
 But as other operators also play some role, significant uncertainties in the estimate $\epe$ have been found.
In order to reduce them  it has been suggested in
\cite{Buras:1993dy} to assume that the $\Delta I=1/2$ rule can completely be
explained by the SM dynamics and to determine the matrix elements of the current-current
operators $Q_1$ and $Q_2$ by using the experimental values of the  $K\to\pi\pi$
isospin amplitudes $A_0$ and $A_2$. This in turn allowed with the help of isospin symmetry to determine  the
matrix elements of $(V-A)\times (V-A)$ operators like $Q_4,Q_9,Q_{10}$.
This method has been used in all analyses performed in Munich since then with
some refinements made recently in \cite{Buras:2015yba,Aebischer:2019mtr}. In this manner the terms like $-2.9$ and $+2.0$ in (\ref{AN2020}) could be  more accurately determined than the remaining dominant terms.

However, this procedure has one
weak point. If NP is responsible for a fraction of the $\Delta I=1/2$ rule
this trick will not provide fully correct results for hadronic matrix elements  and it is safer
to calculate them using  non-perturbative methods like
LQCD or DQCD 
which avoids automatically any NP contributions to hadronic operator  matrix elements. NP can only affect the Wilson coefficients of these operators.
Therefore, in fact the formula (\ref{AN2020}) does not assume that the SM
explains fully the  $\Delta I=1/2$ rule and to find the terms
$-2.9$ and $+2.0$ only hadronic matrix elements from RBC-UKQCD collaboration \cite{Abbott:2020hxn}
have been used, properly evolved to $\mu=1~\gev$. The formula (\ref{AN2020})
exhibits only central values of various contributions. We will return to uncertainties in the final sections of this writeup.

Unfortunately in the 1990s there were no results from LQCD for any of the terms in (\ref{AN2020}) and the relevant hadronic matrix elements
of $Q_6$ and $Q_8$ operators, known only from DQCD at that time, and given by \cite{Buras:1985yx,Buras:1987wc} 
\begin{align}
  \label{eq:Q60}
  \langle Q_6(\mu) \rangle_0 &
  = -\,4  \left[\frac{m_K^2}{m_s(\mu) + m_d(\mu)}\right]^2 (F_K - F_\pi)
    \,\bsi(\mu) ,
\\
  \label{eq:Q82}
  \langle Q_8(\mu) \rangle_2 &
  = \sqrt{2} 
  \left[ \frac{m_K^2}{m_s(\mu) + m_d(\mu)} \right]^2 F_\pi \,\bei(\mu) ,
\end{align}
with $\bsi=\bei=1.0$ in the large N-limit,
were subject
to the large uncertainty in the strange quark mass.
We did not
expose this dependence in (\ref{AN2020}) because in 2021 this uncertainty
is very small.
It is exhibited in the older Munich papers like \cite{Buras:1996dq,Buras:2003zz}. Moreover, in the 1990s it was not clear what was the error on $\bsi$ and $\bei$ so
that often in the phenomenological analyses of the Munich and also Rome
group \cite{Ciuchini:1995cd} their values were varied
within the range $0.8-1.2$.  We will see later that while presently the
$\bei$ is already accurately known, this is not the case for $\bsi$.

In any case it turned out that NLO corrections led to additional suppression of $\epe$ beyond the one by EWP so that typical
values of $\epe$ in Munich papers
  were in the ballpark of $7\times 10^{-4}$
and the ones from Rome more like $3\times 10^{-4}$, both values being
significantly below the experimental result in (\ref{EXP}).

The origin of the difference in $\epe$ resulting from the phenomenological analyses performed in Munich and Rome has been clarified only in 1998 in
a paper written in collaboration with Paolo Gambino and Ulrich Haisch \cite{Buras:1999st}. We have pointed out that without NNLO QCD corrections to EWP contribution the results for $\epe$ are renormalization-scheme dependent and exhibit significant non-physical dependences on the scale $\mu_t$ at which the top-quark mass $m_t(\mu_t)$ is evaluated as well as on the scale $\mu_W$, at which the
full SM is matched onto  the low-energy  effective theory without top, $W^\pm$, $Z$ and
the Higgs. 
It turned out then that the dominant difference between Munich and Rome
NLO results originated in the use of different schemes for $\gamma_5$, NDR
scheme in Munich and t'Hooft-Veltman scheme in Rome. This difference,
as well as scale uncertainties mentioned above,
have been significantly reduced in our paper,  the first NNLO analysis of $\epe$ that concentrated on the EWP contributions. The corresponding
NNLO analysis for QCDPs should hopefully be completed  in 2021 \cite{Cerda-Sevilla:2016yzo,Cerda-Sevilla:2018hjk}, although the
first steps have already been made in 2004 by Martin Gorbahn and Ulrich Haisch,
who calculated the three-loop anomalous dimension matrix of the relevant operators \cite{Gorbahn:2004my}. But we know already today from preliminary results in
 \cite{Cerda-Sevilla:2016yzo,Cerda-Sevilla:2018hjk},
 that while the NNLO QCD corrections to EWPs increase their importance
 in $\epe$  \cite{Buras:1999st}, the ones to QCDPs suppress
the latter. Consequently there is a visible suppression of $\epe$ by the NNLO
QCD corrections. But the percentual effect of this suppression strongly depends
on the values of the relevant hadronic matrix elements which in the case of
QCDPs are still poorly known.

One of the last papers of this period was the one in \cite{Bosch:1999wr} in
which we addressed the first messages from the new round of $\epe$-experiments by NA48 and KTeV collaborations that led eventually to the result in (\ref{EXP}). We found that
it was not possible to explain the data within the SM which motivated a detailed analysis in the MSSM \cite{Buras:2000qz}.

\boldmath
\section{The Third Period: $2000-2015$}\label{P3}
\unboldmath
Already in the second period, but with even increased energy in the third period,
the authors in \cite{Antonelli:1995gw,Bertolini:1995tp,Pallante:1999qf,Pallante:2000hk,Buchler:2001np,Buchler:2001nm,Pallante:2001he}  using 
the ideas from Chiral Perturbation Theory (ChPT) made a strong claim 
that final state interactions (FSI)  enhance $\bsi$ above unity and suppress 
$\bei$ below it so that the SM value for $\epe$ according to them 
is fully consistent with experiment.
Albeit only a very inaccurate value $(17\pm9)\times 10^{-4}$ \cite{Pallante:2001he}   could be obtained at that time. 
More about it later. Critical remarks about some of these  papers appeared in \cite{Buras:2000kx} but they did not slow down the ChPT experts.

The main  theoretical activity on $\epe$ within the SM in this period was another look at the parameter $\hat\Omega_{\rm eff}$ which includes both isospin breaking corrections and QED corrections beyond EWP contributions. Intensive analyses of these corrections can be found in \cite{Cirigliano:2003nn,Cirigliano:2003gt,Bijnens:2004ai} were references to earlier analyses can be found. The renaissance of such analyses with ChPT updates and also including the update of
my 1987 analysis with Jean-Marc G{\'e}rard in \cite{Buras:1987wc} will be described in the fourth period. It modified significantly ChPT results, basically
confirming with smaller uncertainties our 1987 value for $\hat\Omega_{\rm eff}$ that was only slightly lower than  $\hat\Omega_{\rm eff}\approx 0.29$ 
used in our examples above.

In the absence of LQCD results and the claim of ChPT experts that the SM
can reproduce the experimental value in (\ref{EXP}), the interest in $\epe$
among theorists decreased significantly in this period. This was also due to the fact that the new experimental results in other sectors of particle physics, in particular in $B$ physics and neutrino physics, could be easier addressed theoretically than $\epe$. Selected reviews of $\epe$ in this period can be found in
\cite{Bertolini:1998vd,Buras:2003zz,Cirigliano:2011ny,Bertolini:2012pu}. 
On my part I have performed with my PhD students, postdocs and other collaborators  a number of analyses of $\epe$ in various extensions of the SM that
can easily be found in the hep-arXiv. I will not review them here because the input parameters and often the scale of NP changed in the last decade and it is
better to summarize the status of NP models in the last period presented here.
But the formulae presented in these papers could still turn out to be useful one day when the  hadronic uncertainties will be reduced by much.
\boldmath
\section{The Fourth Period: $2015-2019$}\label{P4}
\unboldmath
The new era for $\epe$ began in 2015, when the RBC-UKQCD collaboration  \cite{Bai:2015nea,Blum:2015ywa} presented their first results for $K\to\pi\pi$ hadronic matrix 
elements. From these results 
one could extract values of $\bsi$ and $\bei$ for $\mu=1.30\gev$. They 
 were  \cite{Buras:2015yba}
\be\label{B8LAT0}
\bsi=0.57\pm 0.19,\qquad B_8^{(3/2)}=0.76 \pm 0.05\,,\qquad  {\rm (RBC-UKQCD,~2015)}.
\ee

It is evident that with such values there is a strong cancellation
between QCDP and EWP contributions. With $\hat\Omega_{\rm eff}\approx 0.15$, as used in 2015, and NLO QCD corrections one finds values of $\epe$ in the ballpark of $(1-2)\times 10^{-4}$. This means one order of magnitude below the experimental value. But with an error in the ballpark of $5\times 10^{-4}$ one could talk of an $\epe$ anomaly of at most $3\sigma$.
The relevant analyses that extracted some matrix elements from data  by assuming the $\Delta I=1/2$ rule in the SM can be found in
\cite{Buras:2015yba,Kitahara:2016nld}. The RBC-UKQCD lattice collaboration \cite{Blum:2015ywa, Bai:2015nea}, calculating directly hadronic matrix elements of all operators, but not including isospin breaking (IB) effects, found similar result but with an error of
$7\times 10^{-4}$.

Motivated by the RBC-UKQCD results in (\ref{B8LAT0}), 
 Jean-Marc G{\'e}rard and me calculated already in July 2015 $1/N$ corrections to the large $N$ limit in  (\ref{LN}) \cite{Buras:2015xba}. These corrections, loop corrections in the meson theory with a {\em physical cut-off} $\Lambda\approx 0.7\gev$, are the leading non-factorizable corrections to hadronic matrix elements 
of $Q_6$ and $Q_8$. Two main results in this paper are:
\begin{itemize}
\item
  Realization that the large-$N$ result in (\ref{LN})
is not valid at scales $\ord(1\gev$), as
assumed in all papers before, but at much lower scales 
$\ord(m_\pi^2)$. In order to find it out one has to calculate the non-factorizable $1/N$ contributions represented in DQCD by  meson loops with a physical cut-off $\Lambda$, which separates the long distance and short distance contributions\footnote{For a detailed exposition of this point and comparison with ChPT
  which uses dimensional regularization see Section 3 in \cite{Buras:2014maa}.}. In 
the large $N$ limit one cannot determine the scale in $\bsi$ and $\bei$ and 
as for $\mu\ge 1\gev$ the $\mu$ dependence of these parameters is weak \cite{Buras:1993dy}, without
knowing $1/N$ corrections it was useful to neglect this dependence.
\item
Calculation of $\bsi$ and $\bei$ at scales $\ord(1\gev)$ by performing the meson evolution from  the low factorization scale $\ord(m_\pi^2)$ to the physical 
cutoff $\Lambda$ of DQCD with the result
\be\label{bsibei}
\bsi\le 0.6, \qquad \bei=0.80\pm 0.10,\qquad (\text{DQCD}-2015),
\ee
in a very good agreement with the RBC-UKQCD results in (\ref{B8LAT0}). 
\end{itemize}

As already mentioned, for scales above $1\gev$ both parameters decrease very slowly. This was already  known from the 1993 analysis in \cite{Buras:1993dy}, but as seen in Figs.\,11 and 12 of that paper $\bsi$ decreases faster with increasing scale than $\bei$ in accordance with the pattern at low scales found in 2015 by Jean-Marc and 
myself. This can also be shown analytically \cite{Buras:2015xba}.
Unfortunately, not knowing $1/N$ corrections to $\bsi$ and $\bei$ in 1993, both parameters have been set  at $\mu=m_c$ in \cite{Buras:1993dy} to unity, which is clearly wrong.

While we did not exclude the possibility that our bound on $\bsi$ could be violated by $1/N^2$ 
corrections, vector meson contributions and other effects like final state interactions (FSI) not taken by us into account, one should notice that  with only pseudoscalars included in the loops, the cut-off $\Lambda$ has to  be chosen below $1\gev$ so that these 
 omitted effects, even if they would increase 
 $\bsi$, could still be at least partially compensated by the running to 
 higher scales that are explored by lattice QCD. In any case we expected $\bsi$
 at scales $\ord(1\gev)$ to be below unity.

Therefore it appeared to us in 2015 that we could understand the QCD dynamics
behind the LQCD values which was important for the following
reason. There is no other lattice collaboration beyond RBC-UKQCD one calculating $\bsi$ and $\bei$ at present, so that 
in the lattice world the results of the RBC-UKQCD collaboration for $\epe$
could not be tested in 2015 and this is also the case now.
 As we will emphasize   below, ChPT by itself, has no means 
 to verify or disprove the RBC-UKQCD results for $\bsi$ and $\bei$. As already
 stated above, according to our analysis in \cite{Buras:2015xba},
the main QCD dynamics behind the lattice values in (\ref{B8LAT0}) was the
meson evolution at long distances, analogous to the well known quark-gluon evolution at short distance scales.

At a flavour workshop in Mainz in January 2016 two important ChPT experts, 
Gilberto Colangelo and Antonio Pich, expressed serious doubts about the RBC-UKQCD result in (\ref{B8LAT0}), because the $(\pi\pi)_{I=0}$ 
 phase shift $\delta_0\approx (24\pm 5)^\circ$ obtained by RBC-UKQCD disagreed 
 with  $\delta_0\approx 34^\circ$ obtained by combining dispersion 
theory with experimental input \cite{Colangelo:2001df}.

This criticism appeared 
in print in 2017 \cite{Gisbert:2017vvj} and in two subsequent  conference proceedings \cite{Gisbert:2018tuf,Gisbert:2018niu}. 
It is in line with the one expressed many years ago in \cite{Antonelli:1995gw,Bertolini:1995tp,Pallante:1999qf,Pallante:2000hk,Buchler:2001np,Buchler:2001nm,Pallante:2001he}, but one should realize that with $\delta_0\approx 24^\circ$
a big portion of FSI has been already taken into account  in (\ref{B8LAT0}). Therefore, from Jean-Marc's and my point  of view it appeared rather unlikely
that increasing $\delta_0$ up to its dispersive  value 
$\delta_0\approx 34^\circ$ would shift $\epe$ upwards by one order of magnitude.
In fact, soon after the Mainz workshop we have expressed this view in \cite{Buras:2016fys} demonstrating
that possible enhancement of $\epe$ by FSI cannot be as large as its
suppression through meson evolution absent in the calculations in  \cite{Gisbert:2017vvj}. Subsequently in a Christmas story \cite{Buras:2018ozh} I have illustrated possible impact of the meson evolution on the result of the  authors of
\cite{Gisbert:2018tuf,Gisbert:2018niu},
finding much lower values of $\epe$ than claimed by them. Moreover, based on the insight from DQCD and NNLO QCD corrections
as well as isospin breaking effects, I summarized my expectations for 2026, the 50th anniversary of the first $\epe$ calculation and also my 80th birthday,
by
\be\label{AJBFINAL}
 \boxed{(\epe)_{\rm SM}= (5\pm2)\cdot 10^{-4},\qquad (2026).}
 \ee
 This expectation is in accord with our analysis \cite{Buras:2016fys} from
 which values of $\bsi$ above unity at $\mu=1\gev$ are rather unlikely even after
 the inclusion of FSI. But for $\bsi=1$ we found already (\ref{42}) and
 the slight decrease of $\bei$ could increase the value of $\epe$ by a bit. The
 error is just a guess estimate but even if it is large, confirmation of this result
 by LQCD would imply a significant anomaly and NP at work.

While waiting for the new RBC-UKQCD result, the calculation
of isospin breaking affects and QED corrections, represented by
$\hat\Omega_{\rm eff}$, has been updated within ChPT in 
 \cite{Cirigliano:2019cpi} with the result
\be\label{6.9}
  \hat\Omega_{\rm eff}^{(8)}=(17\pm9)\, 10^{-2}, \qquad (\text{ChPT}-2019),
  \ee
  where the index ``(8)'' indicates that only contributions from the octet
  of pseudoscalars have been taken explicitly into account. The large error of $50\%$ in this estimate originates from the difficulties in the matching of the long distance (LD)
  and short distance (SD) contributions in this framework so that the effects
  of the flavour singlet $\eta_0$ cannot be explicitly included in this
  approach. They are buried in a poorly known low-enery constant $L_7$.

  Using (\ref{6.9}) the most recent 
estimate from ChPT \cite{Cirigliano:2019ani,Gisbert:2020wkb} reads
  \begin{align}
  \label{Pich}
  (\epe)_\text{SM}  & = (14 \pm 5) \times 10^{-4} \,,\qquad (\text{ChPT}-2019).
\end{align}
  It is in contrast to (\ref{AJBFINAL}) fully consistent with experiment
  but because of the large error related to the problematic matching
  of LD and SD contributions in this approach, it still allows for significant
  NP contributions. We will return to this result in the next section.

Parallel to these analytic developments
RBC-UKQCD presented already in 2018 a new 
result for $\delta_0$ that 
with $\delta_0=(32.3\pm 2.1)^\circ$ is within $1\sigma$ from its dispersive
value\footnote{See talks by Ch.Kelly and T. Wang at Lattice 2018.}. 
But the most important recent result from this collaboration is the new
result for $\epe$ to be presented soon.

    The hints for the possible $\epe$ anomaly motivated several authors to perform BSM
  analyses of this ratio. We collected a selection of these papers in Table~\ref{eprimeanomaly} in Appendix~\ref{OPE}. Here we just mention that if this anomaly will be confirmed one day by more
  precise calculations, the leptoquark models, with the
  possible exception of the vector $\text{U}_1$ model, will not be able to
   explain it
 because of the constraints from rare Kaon decays
\cite{Bobeth:2017ecx}. This shows how crucial correlations of $\epe$ with other
observables in a given NP scenario are.  As indicated in Table~\ref{eprimeanomaly}, such correlations  have been analyzed in other NP scenarios.

Moreover, the lessons gained from the SMEFT analysis in \cite{Aebischer:2018csl}
should be very helpful in identifying NP behind possible  $\epe$ anomaly. Such
a general analysis allows to take the constraints from other
processes, in particular from electroweak precision tests and collider
processes,
into account. To this end the master formula for $\epe$ in \cite{Aebischer:2018quc},
valid in any extension of the SM, should facilitate the search for the dynamics
behind the possible anomaly in question. This formula is based on hadronic matrix elements of SM operators from LQCD and the BSM four-quark operators from DQCD
\cite{Aebischer:2018rrz}. The ones of chromomagnetic operators are known
from LQCD \cite{Constantinou:2017sgv} and DQCD \cite{Buras:2018evv}.
They turned out to be less important than expected in the past. This master
formula has been updated in \cite{Aebischer:2020jto} and will be soon generalized to include NLO QCD corrections to the BSM contributions.

\boldmath
\section{The Fifth Period: $2020$}\label{P5}
\unboldmath
The fifth period begins with a surprising new result from 
the RBC-UKQCD collaboration
\cite{Abbott:2020hxn}
\begin{align}
  \label{RBCUKQCD}
  (\epe)_{\rm SM} &
  = (21.7 \pm 8.4) \times 10^{-4} \,,
\end{align}
where statistical, parametric and systematic uncertainties have been added in
quadrature. The central value is by an order of magnitude larger than
the central 2015 value presented by this collaboration but is subject
to large systematic uncertainties which dominate the quoted error. It 
is based on the improved values of the hadronic matrix elements of QCDP,
includes the Wilson coefficients at the NLO level but does not
include  isospin breaking effects and NNLO QCD effects
  
However, as already demonstrated in \cite{Aebischer:2019mtr}, the inclusion
of the effects in question, that are absent in (\ref{RBCUKQCD}) is important.
Indeed after including isospin-breaking
effects from \cite{Cirigliano:2019cpi} in (\ref{6.9})
and NNLO QCD corrections to EWP
contributions \cite{Buras:1999st}, one finds using the hadronic matrix elements of RBC-UKQCD \cite{Aebischer:2020jto}
\begin{align}
  \label{ABBG}
  (\epe)^{(8)}_\text{SM}
  = (17.4 \pm 6.1) \times 10^{-4} \,
\end{align}
instead of (\ref{RBCUKQCD}). The index ``(8)'' indicates that only the octet of pseudoscalars has been included in the evaluation of isospin breaking effects summarized by $\hat\Omega_{\rm eff}^{(8)}$ in (\ref{6.9}).

Yet, already in 1987
  Jean-Marc and me \cite{Buras:1987wc}\footnote{See also \cite{Donoghue:1986nm}.} pointed out that the contribution of $\eta_0$
  and of the resulting $\eta-\eta^\prime$ mixing cannot be neglected in
  the evaluation of $\epe$. Updating
  and significantly improving our 1987 analysis we presented  last spring  the improved estimate of $\hat\Omega_{\rm eff}$  \cite{Buras:2020pjp}
\be\label{6.8}
  \boxed{\hat\Omega_{\rm eff}^{(9)}=(29\pm7)\, 10^{-2},\qquad (\text{NIB}-2020),}
  \ee
  where the index ``(9)'' indicates that the full nonet of pseudoscalars
  has been taken into account. This is the value we have used in our
  examples. Note that the percentual error, even if sizable, amounts to $24\%$,
  a factor of two smaller than in (\ref{6.9}). Most importantly the central
  value is by more than $50\%$ higher than in (\ref{6.9}).

  At the FCPC 2020 conference Toni Pich referred to our result in (\ref{6.8})
  as the {\em naive} IB. As a coauthor of the naive dimensional regularization scheme
  (NDR) \cite{Buras:1989xd} I have no problem with this terminology
  and accepted this name  in (\ref{6.8}). It is well known that the NDR scheme for
  $\gamma_5$ is used these days by almost everybody even if it is not as
  sophisticated as the 't Hooft-Veltman scheme. I expect that the future
  of NIB will be similar. The point is that our analysis in \cite{Buras:2020pjp}
  is really not as naive as one would conclude from its name. In the
  decoupling limit for $\eta_0$ our approach reproduces IB from
  ChPT within $10\%$, but in contrast to ChPT we are able to
  include the effect of $\eta_0$ that is very important. We should
  stress again that in our approach this effect is included explicitly, while
  in the octet scheme, necessarily used in ChPT, it is buried in a poorly
  known low-energy constant $L_7$. But $L_7$ can only be extracted from
  the data in the large $N$-limit which in decays like $K\to\pi\pi$ is
  a bad approximation. Therefore I expect that including $\eta_0$
  effect in $\epe$ will remain a big challenge for ChPT for 
  some time  and whether ChPT will  ever be able to match the present NIB result
  is unclear to me at present.

  Using the value of $\hat\Omega_{\rm eff}^{(9)}$ in (\ref{6.8}) together with
  hadronic matrix elements of RBC-UKQCD one finds
  \cite{Buras:2020pjp,Aebischer:2020jto}
\begin{align}
  \label{BG20}
  \boxed{
  (\epe)^{(9)}_\text{SM}
    = (13.9 \pm 5.2) \times 10^{-4} \,.}
\end{align}
 This is in my view  the present best estimate of this ratio in the SM if one
accepts the the results from RBC-UKQCD on hadronic matrix elements.
However, I will stress below that I am not ready to do it at present
and the main reason for the analyses in  \cite{Buras:2020pjp,Aebischer:2020jto}
was the service to the community by improving the RBC-UKQCD analysis through the addition of  isospin breaking effects
and NNLO QCD corrections to EWPs.

The result in (\ref{BG20}) agrees  well with experiment and with the ChPT expectations (\ref{Pich})
but in view of our comments on the ChPT analysis it is on a more solid footing.
Moreover, as we will demonstrate soon, its agreement with the ChPT value in (\ref{Pich}) is a pure numerical coincidence. We expect further reduction of $\epe$  by roughly $(5-10)\%$ when NNLO QCD
corrections to QCDP contributions will be taken into account
\cite{Cerda-Sevilla:2016yzo, Cerda-Sevilla:2018hjk}. We look forward
to the final results of these authors.

It is a good place to list the values of $\bsi$ and $\bei$ at $\mu=1\gev$
that can be extracted from the most recent RBC-UKQCD results for hadronic matrix elements  for QCDPs \cite{Abbott:2020hxn} and from \cite{Bai:2015nea} for EWPs. This collaboration
calculated them respectively for $\mu=4\gev$ and $\mu=3\gev$ and one
extracts at these scales \cite{Aebischer:2020jto}
\begin{equation}
  \label{eq:Lbsi}
  \bsi(4\gev)   = 1.11 \pm 0.20, \qquad \bei(3\gev) 
    = 0.70 \pm 0.04.
\end{equation}

Performing the RG evolution down to $1\gev$ one finds \cite{Aebischer:2020jto}
instead
\begin{equation}\label{LATB8}
   \bsi(1\gev)   = 1.49 \pm 0.25,  \qquad
  \bei(1\gev) 
    = 0.85 \pm 0.05\,,
\end{equation}
demonstrating the decrease of both parameters with increasing $\mu$ in accordance with DQCD \cite{Buras:2015xba}.

Let us begin  with the good news. Comparing the LQCD value for
$\bei$ with DQCD one in (\ref{bsibei}) we find 
a very good agreement between
LQCD and DQCD as far as EWP contribution to $\epe$ is concerned. This implies
that this contribution to $\epe$, that is unaffected by leading IB corrections, is already
known within the SM with acceptable accuracy:
\be
  \label{EWPSM}
  \boxed{(\epe)^{\text{EWP}}_\text{SM}   = - (7 \pm 1) \times 10^{-4} \,,\qquad (\text{LQCD~and~DQCD}).}
\ee
Because both LQCD and DQCD can perform much better in the case of EWPs than in the case of 
QCDPs I expect that this result will remain with us for coming years.

On the other hand ChPT expected $\bei\approx 0.55$ \cite{Cirigliano:2019ani} with the suppression below unity caused by FSI. Evidently this large suppression has not
been confirmed by the RBC-UKQCD collaboration. Including only the last two terms
in (\ref{AN2020}) we find the EWP contribution estimated by ChPT to be
roughly by a factor of 2 below the result in (\ref{EWPSM}). This already
signals that the agreement of (\ref{BG20}) with (\ref{Pich}) is an
accidental numerical coincidence. This is undermined by the fact that the ChPT
result was obtained with
$\hat\Omega^{(8)}_{\rm eff}$ in place of $\hat\Omega^{(9)}_{\rm eff}$ and 
$\im \lambda_t=(1.35)\times10^{-4}$ instead of our $\im \lambda_t=(1.45)\times10^{-4}$.

The case of QCDPs is a different story. Here the LQCD value  overshoots the DQCD
one by more than a factor of two 
and consequently despite the agreement on EWP contribution the result in (\ref{BG20})
based on RBC-UKQCD hadronic matrix elements differs by roughly a factor of three
from my expectations for 2026 in (\ref{AJBFINAL}). The difference from the
RBC-UKQCD result that does not include IB, QED corrections and NNLO QCD effects in (\ref{RBCUKQCD}) is even by a factor of four. On the other hand
the ChPT estimate of $\bsi$ being in the ballpark of $1.35$ is  at first sight
in the ballpark of the LQCD value. However, such a direct comparison is incorrect because the ChPT value of $\bsi$ corresponds to much lower scales and as demonstrated
in Fig.~2 of \cite{Buras:2018ozh} the inclusion of the meson evolution would
make it significantly smaller, in the ballpark of unity.

Let me next make a few  additional critical remarks about ChPT estimate of $\epe$:
\begin{itemize}
\item
  First of all one should realize that strictly speaking $\epe$ cannot be calculated in ChPT by itself because several important contributions in this framework depend on low-energy constants which have to be taken from LQCD calculations or low energy data.
  While in the case of semi-leptonic decays this procedure is rather successful,
  in the case of $K\to\pi\pi$ it encounters serious problems which I doubt will
  be solved in the coming years.
\item
  In particular the parameters $\bsi$ and $\bei$ are evaluated in the strict
  large $N$ limit which as we stressed corresponds to the scales much lower
  than the scales at which Wilson coefficients can be calculated. But ChPT
  by itself does not have meson evolution and consequently the matching
  with Wilson coefficients is practically impossible. The authors of  \cite{Cirigliano:2019ani}
  admit this stating by themselves that the dominant error in their estimate of $\epe$   originates from their ignorance about  this matching. The low-energy constant which they have to know to overcome this difficulty is $L_5$. Its value obtained from LQCD is still rather uncertain implying large error in (\ref{Pich}).
  However, this error could be an underestimate for the following reasons.
\item
  The expression for $\epe$ in terms of low-energy constants presented by these authors is obtained in the large $N$ limit. While the numerical value for $L_5$
  from LQCD certainly includes some $1/N$ corrections, there could still be  some   missing $1/N$ contributions both in the formula for $\epe$ used by them and also in the
  extraction of $L_5$ in case they would try to find its value from some low-energy data.
\item
  Equally problematic is the inclusion of the singlet $\eta_0$ and of the
  $\eta-\eta^\prime$ mixing which has been known for more than 30 years to suppress $\epe$ significantly \cite{Donoghue:1986nm,Buras:1987wc,Buras:2020pjp}.
  This effect is buried again in a poorly known low energy constant, this time  $L_7$. If the authors of  \cite{Cirigliano:2019ani} would use the NIB of (\ref{6.8}) they would end up, even without the meson evolution, with the value of $\epe$ in the ballpark of $(9-11)\times 10^{-4}$.
\item
  There is one point where ChPT could be superior to DQCD. These are FSI. However, at least in the case of EWP contribution to $\epe$, its strong suppression through FSI predicted by ChPT experts has not been confirmed by RBC-UKQCD which   obtains the result for $\bei$ in agreement with DQCD. Here presently these   effects are not included and the modest suppression of $\bei$ below
  unity in DQCD is due to the meson evolution.
  \end{itemize}

Next let me move to make several comments on the most recent analysis
of the RBC-UKQCD collaboration. I am doing it not only because
all my recent analyses of $\epe$ \cite{Buras:2018ozh,Aebischer:2019mtr,Buras:2019vik,Aebischer:2020jto,Buras:2020pjp} have been simply ignored
by this collaboration.
In particular I want to 
list the  arguments while I still expect the final result for $\epe$ in the SM to be close
to my 2026 expectations and certainly by a factor of at least two below the
present RBC-UKQCD value in (\ref{RBCUKQCD}).

While one should admire the RBC-UKQCD collaboration for their heroic efforts
over at least one decade to calculate
  $\epe$, as they emphasize from first principles, here are my main problems with accepting their result in (\ref{RBCUKQCD}) despite the   large error they admit.
  \begin{itemize}
  \item
    {\em The inclusion of isospin breaking corrections as a symmetric error to their
    value without these corrections.} This is like stating that these corrections
    could in principle enhance $\epe$ implying thereby an anomaly in this ratio
    that would require NP to {\em suppress} this ratio to agree with data.
    Equivalently, it amounts to question all the work done over more than
    thirty years by different authors, with different methods that imply
    significant suppression of $\epe$ by these corrections, in particular by
    the presence of the $\eta_0$ and related $\eta-\eta^\prime$ mixing.
    This is evident by comparing the value  in (\ref{RBCUKQCD}) with
    (\ref{BG20}) which use the same hadronic matrix elements from RBC-UKQCD
    but in  (\ref{BG20}) also isospin breaking corrections are included. While
    it could be legitimate to use LQCD for the calculation of all effects, talking to
    various colleagues who are closer to LQCD than me it is likely that we will not
    see a value for $\epe$ from LQCD including isospin corrections,
    in particular those from $\eta-\eta^\prime$ mixing, before 2026. I do
    hope very much that these expectations are wrong and RBC-UKQCD collaboration or other LQCD groups will surprize us again by calculating this time  these
    corrections with respectable precision.
  \item
    {\em The absence of GIM mechanism, the crucial property of the SM and of a great
    relevance for kaon physics.} RBC-UKQCD works at $4\gev$ without the inclusion
    of charm. While this omission is more important for the QCDP contributions
    to the $\Delta I=1/2$ rule (see below), because in the case of $\epe$ GIM mechanism
    is broken already at higher scales by the disparity of top quark and charm quark masses, it is to be expected that the inclusion of charm will play a role also
    for $\epe$. I expect that this issue will be solved before a satisfactory
    inclusion of isospin breaking effects by LQCD in general.
  \item
   {\em Matching of the lattice renormalization scheme to the $\overline{\rm MS}$
    scheme used for the calculation of Wilson coefficients.} In order to
    reduce the errors in the matching, that is known presently at the one-loop
    level, RBC-UKQCD works at $4\gev$, which without charm is problematic but
    indeed could help in improving the accuracy of the matching because of the smaller value of $\alpha_s$ at this scale than at $\mu=1.5\gev$ used by them in 2015. On the other hand  the problem with the omission of charm at this lower scale is smaller.
\end{itemize}

  Finally, I am looking forward to more accurate LQCD results on $\epe$ from
  Japan \cite{Ishizuka:2018qbn}. This could help in resolving the controversy
  described above.

 \boldmath 
 \section{The Present Status of the $\Delta I=1/2$ Rule}\label{Sec6}
 \unboldmath
 I cannot resist to add a few lines about the $\Delta I=1/2$ rule \cite{GellMann:1955jx,GellMann:1957wh} after
 RBC-UKQCD indicated in  \cite{Abbott:2020hxn} that they should be credited
 for the identification of the dynamics behind this rule. This is disappointing,
 because in their first paper \cite{Boyle:2012ys} they were more careful
 about the history of this rule as are the authors of
\cite{Donini:2020qfu,Hernandez:2020tbc}.
 This history is summarized in our 2014 paper
 \cite{Buras:2014maa} and in particular in Section 7.2.3 of my recent book \cite{Buras:2020xsm}.
 From this it is evident that the credit for the identification of the
 basic dynamics behind this rule should go to the authors of \cite{Bardeen:1986vz} who demonstrated that the current-current operators and not QCDP operators\footnote{The fact that QCDPs cannot be important for the $\Delta I=1/2$ rule has
   also been noticed by the authors of \cite{Chivukula:1986du}.}
 as claimed by the Russian masters \cite{Shifman:1975tn} are dominantly responsible for this rule. While the short-distance contributions analyzed in \cite{Altarelli:1974exa,Gaillard:1974nj} provided only a small enhancement
 of the $K\to\pi\pi$ isospin amplitude $A_0$ over $A_2$ one, the continuation
 of this {\em quark evolution} by {\em meson evolution} in the non-perturbative
 region down to very low energy scales within DQCD approach allowed to
 obtain the enhancement of the ratio ${\RE A_0}/{\RE A_2}$ up to $16\pm 2$ from $\sqrt{2}$, in the absence of QCD dynamics, compared with the experimental value of $22.4$. While a number
 of authors suggested different solutions that were published in the 1990s \cite{Pich:1990mw,Jamin:1994sv,Antonelli:1995gw,Pich:1995qp,Bertolini:1997ir,Bertolini:1998vd,Bijnens:1998ee,Hambye:1998sma,Crewther:2013vea}, the recent result from  the RBC-UKQCD collaboration seems to confirm our findings of 1986 although I have the impression that they cannot see it doing purely numerical  work. In itself it is a very important result and the RBC-UKQCD
 collaboration should be congratulated for it. Also recent dissection of the
 $\Delta I=1/2$ rule at large $N$  by other LQCD experts
 \cite{Donini:2020qfu,Hernandez:2020tbc} shows that in this decade we should know whether NP plays any role in this
 rule. Here also the studies from \cite{Ishizuka:2018qbn} will be important.

 As stated above it is dominantly the meson evolution responsible for
 the enhancement of ${\RE A_0}$ and the suppression of ${\RE A_2}$ with respect
 to the values in which QCD interactions are switched off by going to the large $N$ limit\footnote{Only operator $Q_2$ contributes in this limit and the mesons are non-interacting in this limit. See  \cite{Buras:2014maa,Buras:2020xsm} for a detailed presentation.}:
\be\label{LO}
{\rm Re}A_0=3.59\times 10^{-8}\gev ,\quad   {\rm Re}A_2= 2.54\times 10^{-8}\gev~, \quad \frac{\RE A_0}{\RE A_2}=\sqrt{2},
\ee
in plain disagreement with the experimental value of 22.4.
It should be emphasized that the explanation of the  missing enhancement factor of $15.8$ through some dynamics must simultaneously give the correct values for ${\rm Re}A_0$ and  ${\rm Re}A_2$
\cite{Tanabashi:2018oca}:
\be\label{N1}
{\rm Re}A_0= 27.04(1)\times 10^{-8}~\gev.
\quad
\quad {\rm Re}A_2= 1.210(2)   \times 10^{-8}~\gev.
\ee
This means that these dynamics should suppress  ${\rm Re}A_2$ by a factor of $2.1$, not more, and enhance ${\rm Re}A_0$ by a factor of $7.5$. This tells us 
that while the suppression of  ${\rm Re}A_2$  is an important ingredient in 
the $\Delta I=1/2$ rule, it is not the main origin of this rule.
 It is the enhancement of  ${\rm Re}A_0$,  as
already emphasized in \cite{Shifman:1975tn}, even if, in contrast to this paper, as demonstrated in \cite{Bardeen:1986vz}, the
current-current operators are responsible dominantly for this rule and not 
QCD penguins. More details can be found in our papers and in my book.

I am making this point because the RBC-UKQCD collaboration in their papers and talks stressed more the suppression of ${\rm Re}A_2$ and not
the enhancement of ${\rm Re}A_0$ as the {\em major} dynamics
behind this rule. The simple discussion above shows that this is simply not true. But  working numerically
at $4\gev$ and being not able to switch-off QCD interactions in LQCD one simply  cannot see properly what is really going on. Yet, as demonstrated by Jean-Marc and myself in
\cite{Buras:2018lgu}, in the context of BSM $K^0-\bar K^0$ matrix elements,
meson evolution allows to explain analytically the values obtained by various lattice collaborations \cite{Carrasco:2015pra,Jang:2015sla,Garron:2016mva,Boyle:2017skn,Boyle:2017ssm}. Without the meson evolution one would fail to explain some of these matrix elements by factors of three. Therefore we are convinced that this is also the case of $K\to\pi\pi$, that is of the $\Delta I=1/2$ rule and of $\epe$. In other words meson evolution is bound to be
present in RBC-UKQCD calculations and its effects will hopefully be more
visible in $\epe$ in the next round of calculations by this collaboration.

This is also supported by the comparison
of LQCD with DQCD in this context in Section 9 of \cite{Buras:2014maa} where
we use the language of contractions used by LQCD experts.
As already noticed in \cite{Boyle:2012ys} the dominant two contractions
work constructively to enhance ${\rm Re}A_0$ and destructively to suppress
${\rm Re}A_2$, but at $4\gev$ this only describes in the lattice language
the property of asymptotic freedom that has been found in corresponding
Wilson coefficients by Altarelli and Maiani \cite{Altarelli:1974exa} and Gaillard and Lee \cite{Gaillard:1974nj} by now 45 years ago. This in itself is an
important result because it indicates that a proper matching of hadronic
matrix elements from LQCD with the Wilson coefficients is possible.

However, in order to understand  why the contribution of the sum
of these contractions $2C_1+C_2$ to ${\rm Re}A_0$\footnote{As seen in equation (122)  of \cite{Buras:2014maa} and also in  \cite{Boyle:2012ys} there is a second constructive contribution $C_1+2C_2$ but with a much smaller coefficient multiplying it, it is subleading.} is by a factor of 22 larger than their difference ($C_1-C_2$) entering  ${\rm Re}A_2$, one has to understand physically the dynamics behind the values of $C_1$ and $C_2$. But these
dynamics must be  below $1\gev$, not at $4\gev$. Otherwise the authors of
\cite{Altarelli:1974exa,Gaillard:1974nj} working above $1\gev$ would identify it 45 years ago, but they did not. Thus, while DQCD working below $1\gev$ can
identify these dynamics as the meson evolution, LQCD, working at $4\gev$
can only see its implications in their numerical values which summarize
all the contributions below $4\gev$.

Yet, we agree with CHPT experts that a part of the ${\RE A_0}$ enhancement over
${\RE A_2}$ comes from FSI, although they were not the first to point it out.
To my knowledge these were nuclear physicists \cite{Brown:1990tt} motivated
by our first results in the 1980s.

Finally, let me emphasize the following  problem in the present calculation of the
$\Delta I=1/2$ rule by the RBC-UKQCD collaboration despite the impressive
result they obtained for the ratio in question 
\be\label{EB}
\frac{\RE A_0}{\RE A_2}=19.9(2.3)(4.4), \qquad {\rm RBC-UKQCD~~(2020)}
\ee
that is consistent with the DQCD value of $16\pm2$ and is in  agreement
with the experimental value $22.4$.

The inspection of various contributions to ${\RE A_0}$ in the RBC-UKQCD
  result reveals the surprizing fact
  that at the scale of $4\gev$ the QCDP contribution, known to be governed due
  to GIM mechanism by the $m_c-m_u$ difference,
not only amounts to $10\%$ of ${\RE A_0}$ at this scale but even {\em suppresses} this   amplitude thereby working against the $\Delta I=1/2$ rule. This contribution would   be absent in the presence of charm contributions due to GIM mechanism     shifting the result in (\ref{EB}) towards the experimental value. But in the   presence of charm other contributions could modify this result and
  we have to wait until next round of LQCD calculations which hopefully will include
  GIM mechanism. I am well aware of the fact that QCDP contribution to ${\RE A_0}$ is scale dependent and increases with lowering the scale but
  at $4\gev$ it should be negligible. From DQCD estimates at $1\gev$ it could
  amount to a $10-15\%$ effect but then it would slightly enhance and not suppress ${\RE A_0}$.

  I do hope very much that the RBC-UKQCD collaboration appreciates my
  detailed analysis of their results\footnote{Partly presented already
    in \cite{Aebischer:2020jto,Buras:2020pjp}.}
    despite my reservations and that my critical remarks will motivate them to provide next time much improved results for both $\epe$
  and the $\Delta I=1/2$ rule with all important effects taken into account and with much smaller errors. There is no doubt that from present perspective LQCD will one day give us most precise values for the  $\Delta I=1/2$ rule and $\epe$ within the SM,
  hopefully revealing some NP contributions to both. Yet, this could
  still take a decade of tedious calculations. I expect that other LQCD collaborations like the ones in  \cite{Donini:2020qfu,Ishizuka:2018qbn}  will make similar efforts.

  In summary, in my view the status of the $\Delta I=1/2$ rule as of 2020 is as follows:
  \begin{itemize}
  \item
    The dominant dynamics behind this rule is our beloved QCD. It is simply the
    {\em quark evolution} from $M_W$ down to scale $\ord(1\gev)$ as analysed first
    by  Altarelli and Maiani \cite{Altarelli:1974exa} and Gaillard and Lee \cite{Gaillard:1974nj}, followed by the  {\em meson evolution} down to very low scales at which  QCD becomes  a theory of weakly interacting mesons and free theory of mesons in the strict large $N$ limit \cite{'tHooft:1973jz,'tHooft:1974hx,Witten:1979kh,Treiman:1986ep}.
This non-perturbative evolution 
    within  the Dual QCD approach dominates by far the enhancement of
    ${\RE A_0}$ over ${\RE A_2}$ as demonstrated by Bardeen, G{\'e}rard and myself in  \cite{Bardeen:1986vz,Buras:2014maa}.
    \item
      This picture appears to be confirmed in particular by the RBC-UKQCD
      collaboration \cite{Boyle:2012ys,Abbott:2020hxn} when one takes
      the insight from \cite{Buras:2018lgu} into account,
       but also other LQCD collaborations \cite{Donini:2020qfu,Ishizuka:2018qbn} made significant progress  here. Very importantly, from present perspective only LQCD can provide satisfactory
      estimate of the room left for new physics in this rule. Most
      likely it is at most at the level of $20\%$. A detailed analysis in
      \cite{Buras:2014sba} shows that heavy neutral coloured gauge bosons $G^\prime$
      but not $Z^\prime$ could provide such contributions while satisfying all existing constraints.
            \end{itemize}
    
\boldmath
\section{A Strategy for the Coming Years}
\unboldmath
Evidently, there is no question about that the situation with $\epe$ is very unclear at present. Personally, I am truly delighted that my expectations for $\epe$ in the SM, based
on the collaboration with Jean-Marc, are very different from the ones of RBC-UKQCD and of the ChPT experts. There are two reasons for this. First of all if
we all agreed that the SM agrees with data on $\epe$, the future of $\epe$ would be rather boring. Equally important, if one day my expectations in (\ref{AJBFINAL}) will be confirmed by several LQCD collaborations, it will be evident
who should be credited for the identification of the $\epe$-anomaly.

However, even ChPT practitioners and RBC-UKQCD
collaboration, who strongly disagree with Jean-Marc's and my claims
about $\epe$, cannot exclude that at a certain level NP will be required to fit its
experimental value. Yet, analyzing NP models containing  new parameters for various values of $\bsi$ and $\bei$ complicates the search for NP  by much. Here comes
one idea which in my view could give us a clue which NP models could have a chance to explain possible anomaly dependent on its size  \cite{Buras:2015jaq}.

Instead, of varying $\bsi$ and $\bei$ we can just write
\be\label{GENERAL}
\frac{\varepsilon'}{\varepsilon}=\left(\frac{\varepsilon'}{\varepsilon}\right)^{\rm SM}+\left(\frac{\varepsilon'}{\varepsilon}\right)^{\rm BSM}\,
\ee
and assume
that NP provides a  shift in $\epe$:
\be\label{deltaeps}
\left(\frac{\varepsilon'}{\varepsilon}\right)^{\rm BSM}= \kepe\cdot 10^{-3}, \qquad   -0.5 \le \kepe \le 1.0,
\ee
with the range for $\kepe$ indicating conservatively the room left for BSM contributions. This range is dictated by the recent analyses in \cite{Buras:2020pjp,Aebischer:2020jto}. Personally, I would vary $\kepe$ only in the range
$0.5\le \kepe \le 1.5$ but this would mean ignoring the results from ChPT and RBC-UKQCD which I do not want to do today. We are fortunate that there is no interference between these two contributions although the Wilson coefficients of SM
 operators can be affected by NP. The corresponding modifications are included in the BSM term.

Now, in a given NP model $\epe$ is correlated with other observables, in particular those in the K-meson decays like $\klpn$, $\kpn$, $K_{L}\rightarrow\pi^0\mu^+\mu^-$ and $K_{S}\rightarrow\mu^+\mu^-$.
As a consequence
one can study the dependence of the corresponding branching ratios as functions
of $\kepe$  which depends on the model considered. We refer to numerous plots
of such dependences in  \cite{Buras:2015jaq,Aebischer:2020mkv}.

One can now ask what is the uncertainty in BSM contributions due to hadronic
matrix elements. Here comes a good news. It turns out that in most
NP models considered until now the dominant shift in $\epe$ comes from the
modifications of the EWP contributions because similar
to isospin breaking effects they are enhanced by a factor ${\RE A_0/\RE A_2}$
relative to those of QCDPs. But $\bei$ is already well known
so that the main uncertainty in NP contributions comes in this case from new parameters in BSM models which can be constrained by other processes.

Yet,  there are also contributions from BSM operators, in particular scalar
and tensor operators. Their matrix elements have only been calculated within
DQCD \cite{Aebischer:2018rrz}. But the master formula for all BSM scenarios
presented in \cite{Aebischer:2018quc,Aebischer:2018csl} demonstrates very clearly the dominance of the $\Delta I=3/2$
contributions over the $\Delta I=1/2$ ones also in this case. While the computation of the matrix elements of these new operators within LQCD has still to be done, it is expected
that the uncertainties in the dominant $\Delta I=3/2$ contributions will
be smaller than the present uncertainty in the matrix element of the QCDP operator $Q_6$.

\boldmath
\section{Final Remarks}
\unboldmath
Our $\epe$-story approaches the end. I have concentrated here on the non-perturbative calculations because no consensus has been reached among theorists until now. However, being privileged to be one of two\footnote{The second is Guido Martinelli in the context of the contributions of chromomagnetic penguins to $\epe$
\cite{Constantinou:2017sgv}.}   
theorists who calculated both short distance and long distance contributions to $\epe$, I want to emphasize that without NLO calculations of Wilson coefficients performed in the early 1990s in Munich
\cite{Buras:1991jm,Buras:1992tc,Buras:1992zv,Buras:1993dy} and Rome \cite{Ciuchini:1992tj,Ciuchini:1993vr}, the uncertainties in the prediction for $\epe$ would
be even larger. In particular without these corrections the matching of Wilson coefficients to hadronic matrix elements performed by LQCD would not be possible.
The result would be simply renormalization scheme dependent.
The story of these calculations is described  in \cite{Buras:2011we} and in
this context I want to thank Guido Martinelli and his strong team
for a very friendly competition we had.

The uncertainties in various steps leading
to (\ref{BG20}) should still be significantly decreased in the coming years
and I do hope very much that by 2026 the picture of $\epe$ with respect to possible NP contributions will be much clearer than it is today. In particular 
the  identification of new sources of CP violation in the data of
NA48 and KTeV collaborations would be very important because they could play in principle a role in the explanation of our  existence.   I really have no idea whether NP in $\epe$, if found, would be  responsible for our existence. Not only because this is presently beyond my  skills but also because we did not yet identify what this NP could be, although several ideas have been put forward. They are listed in Table~ \ref{eprimeanomaly}. My bet would be a heavy $Z^\prime$ and/or vector-like quarks but
not leptoquarks.

My recollection of the $\epe$ efforts dealt dominantly with theory.
Yet, without the measurements of $\epe$ by NA48 and 
KTeV collaborations, all these discussions between RBC-UKQCD, ChPT and DQCD 
 experts 
would be much less exciting and we should thank these two important experimental groups 
for the result in (\ref{EXP}).

\section{Acknowledgements}
Over 37 years I had 39 collaborators with whom I have
written papers on $\epe$ but Jean-Marc G{\'e}rard is the one with whom
I have written the highest number of $\epe$-papers and in fact among the {\em male} physicists the one 
with whom I have written the highest number of papers to date. In
particular this includes papers on the $\Delta I=1/2$ rule, and large $N$ calculations in the context of the Dual QCD approach, 20 journal papers in total. Only Monika Blanke and Jennifer Girrbach-Noe can compete
with him in this respect. The important virtue of this collaboration was that we had to struggle for 35 years against the spanish matadors, of three physics generations by now,
Eduardo de Rafael, Antonio Pich and Hector Gisbert. In this context we were
declared to be {\em naive}, both as far as final state interactions and isospin breaking corrections to $\epe$ are concerned, which clearly united us. But fortunately the collected joint number of citations (2656)  demonstrates that our work has not been ignored by the community and there is no doubt that this number will increase in the future.  I want
to thank Jean-Marc for this great and most pleasant  collaboration and in particular
for his deep insight into low-energy QCD from which I benefited in many ways.
Thanks go also to those with whom I
performed many analyses of $\epe$ within the SM and in various NP scenarios.
Several of their names appeared in the reference list to this writing. The remaining ones can be found in INSPIRE.

Finally, I would like to thank the Max-Planck Institute for Physics, the Physics Department at TUM, the TUM Institute for Advanced Study, DFG (German Research Foundation), BMFT (Federal Ministry of Education and Research), the Alexander von  Humboldt Foundation, European Research
Council and two Excellence Clusters (2006-2019), founded by the DFG under Germany's Excellence Strategy, for the financial support over 37 years in the research presented
here. For the coming years the main support for my research will come from the TUM-IAS and the Alexander von  Humboldt Foundation. But also support from the Excellence Cluster ORIGINS – EXC-2094 – 390783311 is highly appreciated.

\appendix
\section{Operators and New Physics Analyses}\label{OPE}
 We list the operators mentioned in the text:

{\bf Current--Current:}
\begin{equation}\label{O1s} 
Q_1 = (\bar s_{\alpha} u_{\beta})_{V-A}\;(\bar u_{\beta} d_{\alpha})_{V-A},
~~~~~~Q_2 = (\bar su)_{V-A}\;(\bar ud)_{V-A},
\end{equation}

{\bf QCD Penguins:}
\begin{equation}\label{O2s}
Q_3 = (\bar s d)_{V-A}\!\!\sum_{q=u,d,s,c,b}(\bar qq)_{V-A},~~~~~   
Q_4 = (\bar s_{\alpha} d_{\beta})_{V-A}\!\!\sum_{q=u,d,s,c,b}(\bar q_{\beta} 
       q_{\alpha})_{V-A} ,
\end{equation}
\begin{equation}\label{O3s}
Q_5 = (\bar s d)_{V-A}\!\!\sum_{q=u,d,s,c,b}(\bar qq)_{V+A},~~~~~
Q_6 = (\bar s_{\alpha} d_{\beta})_{V-A}\!\!\sum_{q=u,d,s,c,b}
      (\bar q_{\beta} q_{\alpha})_{V+A}, 
\end{equation}

{\bf Electroweak Penguins:}
\begin{equation}\label{O4s} 
Q_7 = \frac{3}{2}\,(\bar s d)_{V-A}\!\!\sum_{q=u,d,s,c,b} e_q\,(\bar qq)_{V+A}, 
~~~~~Q_8 = \frac{3}{2}\,(\bar s_{\alpha} d_{\beta})_{V-A}\!\!\sum_{q=u,d,s,c,b}
      e_q\,(\bar q_{\beta} q_{\alpha})_{V+A},
\end{equation}
\begin{equation}\label{O5s} 
 Q_9 = \frac{3}{2}\,(\bar s d)_{V-A}\!\!\sum_{q=u,d,s,c,b}e_q\,(\bar q q)_{V-A},
~~~~~Q_{10} =\frac{3}{2}\,
(\bar s_{\alpha} d_{\beta})_{V-A}\!\!\sum_{q=u,d,s,c,b}e_q\,
       (\bar q_{\beta}q_{\alpha})_{V-A} \,.
\end{equation}
Here, $\alpha,\beta$ denote colour indices and $e_q$ denotes the electric quark
charges reflecting the electroweak origin of $Q_7,\ldots,Q_{10}$. Finally,
$(\bar sd)_{V-A}\equiv \bar s_\alpha\gamma_\mu(1-\gamma_5) d_\alpha$. 

  \begin{table}
\renewcommand{\arraystretch}{1.3}
\centering
\resizebox{\columnwidth}{!}{
\begin{tabular}{|c|c|c|}
\hline
  NP Scenario & References  & Correlations with
\\
\hline\hline
  LHT
& \cite{Blanke:2015wba}
& $\klpn$
\\
  $Z$-FCNC
& \cite{Buras:2015jaq, Bobeth:2017xry, Endo:2016tnu}
& $\kpn$ and $\klpn$
\\
  $Z^\prime$
& \cite{Buras:2015jaq}
& $\kpn$, $\klpn$ and $\Delta M_K$
\\
  Simplified Models
& \cite{Buras:2015yca}
& $\klpn$
\\
  331 Models
& \cite{Buras:2015kwd, Buras:2016dxz}
& $b\to s\ell^+\ell^-$
\\
  Vector-Like Quarks
& \cite{Bobeth:2016llm}
& $\kpn$, $\klpn$ and $\Delta M_K$
\\
  Supersymmetry
& \cite{Tanimoto:2016yfy, Kitahara:2016otd, Endo:2016aws, Crivellin:2017gks, Endo:2017ums}
& $\kpn$ and $\klpn$
\\
  2HDM
& \cite{Chen:2018ytc, Chen:2018vog}
& $\kpn$ and $\klpn$
\\
  Right-handed Currents
& \cite{Cirigliano:2016yhc, Alioli:2017ces}
& EDMs
\\
  Left-Right Symmetry
& \cite{Haba:2018byj, Haba:2018rzf}
&  EDMs
\\
  Leptoquarks
& \cite{Bobeth:2017ecx}
& all rare Kaon decays
\\
  SMEFT
& \cite{Aebischer:2018csl}
& several processes
\\
  $\text{SU(8)}$
& \cite{Matsuzaki:2018jui}
&  $b\to s\ell^+\ell^-$, $\kpn$, $\klpn$
\\
  Diquarks
& \cite{Chen:2018dfc, Chen:2018stt}
& $\varepsilon_K$, $\kpn$, $\klpn$
\\
  3HDM + $\nu_R$
& \cite{Marzo:2019ldg}
&  $R(K^{(*)})$, $R(D^{(*)})$
\\
  Vectorlike compositeness
& \cite{Matsuzaki:2019clv}
&  $R(K^{(*)})$, $R(D^{(*)})$, $\varepsilon_K$, $\kpn$, $\klpn$
\\
  $\text{U(2)}^3$ flavour symmetry
& \cite{Crivellin:2019isj}
& hadronic $B\to K\pi$, $B_{s,d}\to (KK, \pi\pi)$, $B_s\to \phi (\rho^0,\pi^0)$
\\
\hline
\end{tabular}
}
\caption{Papers studying implications of a possible $\epe$ anomaly.
  \label{eprimeanomaly}
}
\end{table}

\renewcommand{\refname}{R\lowercase{eferences}}

\addcontentsline{toc}{section}{References}

\bibliographystyle{JHEP}
\bibliography{Bookallrefs}

\providecommand{\href}[2]{#2}\begingroup\raggedright\begin{thebibliography}{100}

\bibitem{Aebischer:2020jto}
J.~Aebischer, C.~Bobeth, and A.~J. Buras, {\it {$\varepsilon '/\varepsilon $ in
  the Standard Model at the Dawn of the 2020s}},  {\em Eur. Phys. J. C} {\bf
  80} (2020), no.~8 705, [\href{http://arxiv.org/abs/2005.05978}{{\tt
  arXiv:2005.05978}}].

\bibitem{Batley:2002gn}
{\bf NA48} Collaboration, J.~Batley et~al., {\it {A Precision measurement of
  direct CP violation in the decay of neutral kaons into two pions}},  {\em
  Phys.~Lett.} {\bf B544} (2002) 97--112,
  [\href{http://arxiv.org/abs/hep-ex/0208009}{{\tt hep-ex/0208009}}].

\bibitem{AlaviHarati:2002ye}
{\bf KTeV} Collaboration, A.~Alavi-Harati et~al., {\it {Measurements of direct
  CP violation, CPT symmetry, and other parameters in the neutral kaon
  system}},  {\em Phys.~Rev.} {\bf D67} (2003) 012005,
  [\href{http://arxiv.org/abs/hep-ex/0208007}{{\tt hep-ex/0208007}}].

\bibitem{Abouzaid:2010ny}
{\bf KTeV} Collaboration, E.~Abouzaid et~al., {\it {Precise Measurements of
  Direct CP Violation, CPT Symmetry, and Other Parameters in the Neutral Kaon
  System}},  {\em Phys. Rev.} {\bf D83} (2011) 092001,
  [\href{http://arxiv.org/abs/1011.0127}{{\tt arXiv:1011.0127}}].

\bibitem{Ellis:1976fn}
J.~R. Ellis, M.~K. Gaillard, and D.~V. Nanopoulos, {\it {Lefthanded Currents
  and CP Violation}},  {\em Nucl. Phys.} {\bf B109} (1976) 213--243.

\bibitem{Shifman:1975tn}
M.~A. Shifman, A.~Vainshtein, and V.~I. Zakharov, {\it {Light Quarks and the
  Origin of the $\Delta I = 1/2$ Rule in the Nonleptonic Decays of Strange
  Particles}},  {\em Nucl.~Phys.} {\bf B120} (1977) 316.

\bibitem{Altarelli:1974exa}
G.~Altarelli and L.~Maiani, {\it {Octet Enhancement of Nonleptonic Weak
  Interactions in Asymptotically Free Gauge Theories}},  {\em Phys. Lett.} {\bf
  B52} (1974) 351--354.

\bibitem{Gaillard:1974nj}
M.~Gaillard and B.~W. Lee, {\it {$\Delta I = 1/2$ Rule for Nonleptonic Decays
  in Asymptotically Free Field Theories}},  {\em Phys.~Rev.~Lett.} {\bf 33}
  (1974) 108.

\bibitem{Bardeen:1986vz}
W.~A. Bardeen, A.~J. Buras, and J.-M. G\'erard, {\it {A Consistent Analysis of
  the $\Delta I = 1/2$ Rule for K Decays}},  {\em Phys.~Lett.} {\bf B192}
  (1987) 138.

\bibitem{Abbott:2020hxn}
{\bf RBC, UKQCD} Collaboration, R.~Abbott et~al., {\it {Direct CP violation and
  the $\Delta I=1/2$ rule in $K\to\pi\pi$ decay from the standard model}},
  {\em Phys. Rev. D} {\bf 102} (2020), no.~5 054509,
  [\href{http://arxiv.org/abs/2004.09440}{{\tt arXiv:2004.09440}}].

\bibitem{Gilman:1978wm}
F.~J. Gilman and M.~B. Wise, {\it {The $\Delta I = 1/2$ Rule and Violation of
  CP in the Six Quark Model}},  {\em Phys. Lett.} {\bf 83B} (1979) 83--86.

\bibitem{Guberina:1979ix}
B.~Guberina and R.~D. Peccei, {\it {Quantum Chromodynamic Effects and CP
  Violation in the Kobayashi-Maskawa Model}},  {\em Nucl. Phys.} {\bf B163}
  (1980) 289--311.

\bibitem{Gilman:1979bc}
F.~J. Gilman and M.~B. Wise, {\it {Effective Hamiltonian for $\Delta S = 1$
  Weak Nonleptonic Decays in the Six Quark Model}},  {\em Phys. Rev.} {\bf D20}
  (1979) 2392.

\bibitem{Buras:1983ap}
A.~J. Buras, W.~Slominski, and H.~Steger, {\it {B Meson Decay, CP Violation,
  Mixing Angles and the Top Quark Mass}},  {\em Nucl. Phys.} {\bf B238} (1984)
  529--560.

\bibitem{Donoghue:1986nm}
J.~F. Donoghue, E.~Golowich, B.~R. Holstein, and J.~Trampetic, {\it
  {Electromagnetic and Isospin Breaking Effects Decrease $\epe$}},  {\em Phys.
  Lett.} {\bf B179} (1986) 361. [Erratum: Phys. Lett.B188,511(1987)].

\bibitem{Buras:1987wc}
A.~J. Buras and J.~M. G\'erard, {\it {Isospin Breaking Contributions to
  $\epe$}},  {\em Phys.~Lett.} {\bf B192} (1987) 156.

\bibitem{Bijnens:1983ye}
J.~Bijnens and M.~B. Wise, {\it {Electromagnetic Contribution to $\epe$}},
  {\em Phys. Lett.} {\bf B137} (1984) 245--250.

\bibitem{Sharpe:1987cx}
S.~R. Sharpe, {\it {On the Contribution of Electromagnetic Penguins to
  $\epsilon^\prime$}},  {\em Phys. Lett. B} {\bf 194} (1987) 551--556.

\bibitem{Lusignoli:1988fz}
M.~Lusignoli, {\it {Electromagnetic Corrections to the Effective Hamiltonian
  for Strangeness Changing Decays and $\epsilon^\prime / \epsilon$}},  {\em
  Nucl. Phys. B} {\bf 325} (1989) 33--61.

\bibitem{Bardeen:1986vp}
W.~A. Bardeen, A.~J. Buras, and J.-M. G\'erard, {\it {The $\Delta I = 1/2$ Rule
  in the Large $N$ Limit}},  {\em Phys.~Lett.} {\bf B180} (1986) 133.

\bibitem{Bardeen:1986uz}
W.~A. Bardeen, A.~J. Buras, and J.-M. G\'erard, {\it {The $K\to\pi \pi$ Decays
  in the Large N Limit: Quark Evolution}},  {\em Nucl.~Phys.} {\bf B293} (1987)
  787.

\bibitem{Buras:1987qa}
A.~J. Buras and J.~G{\'e}rard, {\it {$\epe$ in the Standard Model}},  {\em
  Phys.~Lett.} {\bf B203} (1988) 272.

\bibitem{Dib:1988md}
C.~Dib, I.~Dunietz, and F.~J. Gilman, {\it {{CP} Violation in the $K_L \to
  \pi^0 \ell^+ \ell^-$ Decay Amplitude for Large $m_t$}},  {\em Phys. Lett.}
  {\bf B218} (1989) 487--492.

\bibitem{Flynn:1988ve}
J.~Flynn and L.~Randall, {\it {The {CP} Violating Contribution to the Decay
  $K_L \to \pi^0 e^+ e^-$}},  {\em Nucl. Phys.} {\bf B326} (1989) 31. [Erratum:
  Nucl. Phys.B334,580(1990)].

\bibitem{Buras:1989xd}
A.~J. Buras and P.~H. Weisz, {\it {QCD Nonleading Corrections to Weak Decays in
  Dimensional Regularization and 't Hooft-Veltman Schemes}},  {\em Nucl. Phys.}
  {\bf B333} (1990) 66--99.

\bibitem{Flynn:1989iu}
J.~M. Flynn and L.~Randall, {\it {The Electromagnetic Penguin Contribution to
  $\varepsilon^\prime / \varepsilon$ for Large Top Quark Mass}},  {\em
  Phys.~Lett.} {\bf B224} (1989) 221.

\bibitem{Buchalla:1989we}
G.~Buchalla, A.~J. Buras, and M.~K. Harlander, {\it {The Anatomy of
  $\varepsilon' / \varepsilon$ in the Standard Model}},  {\em Nucl.~Phys.} {\bf
  B337} (1990) 313--362.

\bibitem{Paschos:1991as}
E.~A. Paschos and Y.~L. Wu, {\it {Correlations between $\epe$ and heavy top}},
  {\em Mod. Phys. Lett.} {\bf A6} (1991) 93--106.

\bibitem{Lusignoli:1991bm}
M.~Lusignoli, L.~Maiani, G.~Martinelli, and L.~Reina, {\it {Mixing and CP
  violation in K and B mesons: A Lattice QCD point of view}},  {\em Nucl.
  Phys.} {\bf B369} (1992) 139--170.

\bibitem{Burkhardt:1988yh}
{\bf NA31} Collaboration, H.~Burkhardt et~al., {\it {First Evidence for Direct
  CP Violation}},  {\em Phys. Lett.} {\bf B206} (1988) 169--176.

\bibitem{Barr:1993rx}
{\bf NA31} Collaboration, G.~D. Barr et~al., {\it {A New measurement of direct
  CP violation in the neutral kaon system}},  {\em Phys. Lett.} {\bf B317}
  (1993) 233--242.

\bibitem{Gibbons:1993zq}
L.~K. Gibbons et~al., {\it {Measurement of the CP violation parameter
  Re($\epsilon^{\prime} / \epsilon$)}},  {\em Phys. Rev. Lett.} {\bf 70} (1993)
  1203--1206.

\bibitem{Buras:1991jm}
A.~J. Buras, M.~Jamin, M.~E. Lautenbacher, and P.~H. Weisz, {\it {Effective
  Hamiltonians for $\Delta S = 1$ and $\Delta B = 1$ nonleptonic decays beyond
  the leading logarithmic approximation}},  {\em Nucl. Phys.} {\bf B370} (1992)
  69--104. [Addendum: Nucl. Phys.B375,501(1992)].

\bibitem{Buras:1992tc}
A.~J. Buras, M.~Jamin, M.~E. Lautenbacher, and P.~H. Weisz, {\it {Two loop
  anomalous dimension matrix for $\Delta S = 1$ weak nonleptonic decays. 1.
  $\ord(\alpha_s^2)$}},  {\em Nucl.~Phys.} {\bf B400} (1993) 37--74,
  [\href{http://arxiv.org/abs/hep-ph/9211304}{{\tt hep-ph/9211304}}].

\bibitem{Buras:1992zv}
A.~J. Buras, M.~Jamin, and M.~E. Lautenbacher, {\it {Two loop anomalous
  dimension matrix for $\Delta S = 1$ weak nonleptonic decays. 2.
  $\ord(\alpha\alpha_s)$}},  {\em Nucl.~Phys.} {\bf B400} (1993) 75--102,
  [\href{http://arxiv.org/abs/hep-ph/9211321}{{\tt hep-ph/9211321}}].

\bibitem{Buras:1993dy}
A.~J. Buras, M.~Jamin, and M.~E. Lautenbacher, {\it {The Anatomy of
  $\varepsilon'/ \varepsilon$ beyond leading logarithms with improved hadronic
  matrix elements}},  {\em Nucl.~Phys.} {\bf B408} (1993) 209--285,
  [\href{http://arxiv.org/abs/hep-ph/9303284}{{\tt hep-ph/9303284}}].

\bibitem{Ciuchini:1992tj}
M.~Ciuchini, E.~Franco, G.~Martinelli, and L.~Reina, {\it {$\epe$ at the
  Next-to-leading order in QCD and QED}},  {\em Phys. Lett.} {\bf B301} (1993)
  263--271, [\href{http://arxiv.org/abs/hep-ph/9212203}{{\tt hep-ph/9212203}}].

\bibitem{Ciuchini:1993vr}
M.~Ciuchini, E.~Franco, G.~Martinelli, and L.~Reina, {\it {The $\Delta S = 1$
  effective Hamiltonian including next-to-leading order QCD and QED
  corrections}},  {\em Nucl. Phys.} {\bf B415} (1994) 403--462,
  [\href{http://arxiv.org/abs/hep-ph/9304257}{{\tt hep-ph/9304257}}].

\bibitem{Buchalla:1995vs}
G.~Buchalla, A.~J. Buras, and M.~E. Lautenbacher, {\it {Weak decays beyond
  leading logarithms}},  {\em Rev.~Mod.~Phys.} {\bf 68} (1996) 1125--1144,
  [\href{http://arxiv.org/abs/hep-ph/9512380}{{\tt hep-ph/9512380}}].

\bibitem{Buras:2011we}
A.~J. Buras, {\it {Climbing NLO and NNLO Summits of Weak Decays}},
  \href{http://arxiv.org/abs/1102.5650}{{\tt arXiv:1102.5650}}.

\bibitem{Buras:2020xsm}
A.~J. Buras, {\em {Gauge Theory of Weak Decays}}.
\newblock Cambridge University Press, 6, 2020.

\bibitem{Buras:1985yx}
A.~J. Buras and J.-M. G\'erard, {\it {$1/N$ Expansion for Kaons}},  {\em
  Nucl.~Phys.} {\bf B264} (1986) 371.

\bibitem{Boyle:2012ys}
{\bf RBC, UKQCD} Collaboration, P.~Boyle et~al., {\it {Emerging understanding
  of the $\Delta I = 1/2$ Rule from Lattice QCD}},  {\em Phys. Rev. Lett.} {\bf
  110} (2013), no.~15 152001, [\href{http://arxiv.org/abs/1212.1474}{{\tt
  arXiv:1212.1474}}].

\bibitem{Buras:2015yba}
A.~J. Buras, M.~Gorbahn, S.~J{\"a}ger, and M.~Jamin, {\it {Improved anatomy of
  $\varepsilon'/\varepsilon$ in the Standard Model}},  {\em JHEP} {\bf 11}
  (2015) 202, [\href{http://arxiv.org/abs/1507.06345}{{\tt arXiv:1507.06345}}].

\bibitem{Aebischer:2019mtr}
J.~Aebischer, C.~Bobeth, and A.~J. Buras, {\it {On the importance of NNLO QCD
  and isospin-breaking corrections in $\varepsilon '/\varepsilon $}},  {\em
  Eur. Phys. J.} {\bf C80} (2020), no.~1 1,
  [\href{http://arxiv.org/abs/1909.05610}{{\tt arXiv:1909.05610}}].

\bibitem{Buras:1996dq}
A.~J. Buras, M.~Jamin, and M.~E. Lautenbacher, {\it {A 1996 analysis of the CP
  violating ratio $\epe$}},  {\em Phys.~Lett.} {\bf B389} (1996) 749--756,
  [\href{http://arxiv.org/abs/hep-ph/9608365}{{\tt hep-ph/9608365}}].

\bibitem{Buras:2003zz}
A.~J. Buras and M.~Jamin, {\it {$\varepsilon'/\varepsilon$ at the NLO: 10 years
  later}},  {\em JHEP} {\bf 01} (2004) 048,
  [\href{http://arxiv.org/abs/hep-ph/0306217}{{\tt hep-ph/0306217}}].

\bibitem{Ciuchini:1995cd}
M.~Ciuchini, E.~Franco, G.~Martinelli, L.~Reina, and L.~Silvestrini, {\it {An
  Upgraded analysis of $\epe$ at the next-to-leading order}},  {\em Z. Phys.}
  {\bf C68} (1995) 239--256, [\href{http://arxiv.org/abs/hep-ph/9501265}{{\tt
  hep-ph/9501265}}].

\bibitem{Buras:1999st}
A.~J. Buras, P.~Gambino, and U.~A. Haisch, {\it {Electroweak penguin
  contributions to non-leptonic $\Delta F = 1$ decays at NNLO}},  {\em
  Nucl.~Phys.} {\bf B570} (2000) 117--154,
  [\href{http://arxiv.org/abs/hep-ph/9911250}{{\tt hep-ph/9911250}}].

\bibitem{Cerda-Sevilla:2016yzo}
M.~Cerd{\'a}-Sevilla, M.~Gorbahn, S.~J{\"a}ger, and A.~Kokulu, {\it {Towards
  NNLO accuracy for $\epe$}},  {\em J. Phys. Conf. Ser.} {\bf 800} (2017),
  no.~1 012008, [\href{http://arxiv.org/abs/1611.08276}{{\tt
  arXiv:1611.08276}}].

\bibitem{Cerda-Sevilla:2018hjk}
M.~Cerdà-Sevilla, {\it {NNLO QCD Contributions to $\varepsilon ^\prime
  /\varepsilon $}},  {\em Acta Phys. Polon.} {\bf B49} (2018) 1087--1096.

\bibitem{Gorbahn:2004my}
M.~Gorbahn and U.~Haisch, {\it {Effective Hamiltonian for non-leptonic $|\Delta
  F| = 1$ decays at NNLO in QCD}},  {\em Nucl.~Phys.} {\bf B713} (2005)
  291--332, [\href{http://arxiv.org/abs/hep-ph/0411071}{{\tt hep-ph/0411071}}].

\bibitem{Bosch:1999wr}
S.~Bosch et~al., {\it Standard model confronting new results for
  $\varepsilon'/\varepsilon$},  {\em Nucl.~Phys.} {\bf B565} (2000) 3--37,
  [\href{http://arxiv.org/abs/hep-ph/9904408}{{\tt hep-ph/9904408}}].

\bibitem{Buras:2000qz}
A.~J. Buras, P.~Gambino, M.~Gorbahn, S.~J{\"a}ger, and L.~Silvestrini, {\it
  {$\varepsilon'/\varepsilon$ and rare $K$ and $B$ decays in the MSSM}},  {\em
  Nucl.~Phys.} {\bf B592} (2001) 55--91,
  [\href{http://arxiv.org/abs/hep-ph/0007313}{{\tt hep-ph/0007313}}].

\bibitem{Antonelli:1995gw}
V.~Antonelli, S.~Bertolini, M.~Fabbrichesi, and E.~I. Lashin, {\it {The $\Delta
  I = 1/2$ selection rule}},  {\em Nucl. Phys.} {\bf B469} (1996) 181--201,
  [\href{http://arxiv.org/abs/hep-ph/9511341}{{\tt hep-ph/9511341}}].

\bibitem{Bertolini:1995tp}
S.~Bertolini, J.~O. Eeg, and M.~Fabbrichesi, {\it {A New estimate of $\epe$}},
  {\em Nucl. Phys.} {\bf B476} (1996) 225--254,
  [\href{http://arxiv.org/abs/hep-ph/9512356}{{\tt hep-ph/9512356}}].

\bibitem{Pallante:1999qf}
E.~Pallante and A.~Pich, {\it {Strong enhancement of $\varepsilon'/\varepsilon$
  through final state interactions}},  {\em Phys. Rev. Lett.} {\bf 84} (2000)
  2568--2571, [\href{http://arxiv.org/abs/hep-ph/9911233}{{\tt
  hep-ph/9911233}}].

\bibitem{Pallante:2000hk}
E.~Pallante and A.~Pich, {\it {Final state interactions in kaon decays}},  {\em
  Nucl. Phys.} {\bf B592} (2001) 294--320,
  [\href{http://arxiv.org/abs/hep-ph/0007208}{{\tt hep-ph/0007208}}].

\bibitem{Buchler:2001np}
M.~B{\"u}chler, G.~Colangelo, J.~Kambor, and F.~Orellana, {\it {A Note on the
  dispersive treatment of $K\to\pi\pi$ with the kaon off-shell}},  {\em Phys.
  Lett.} {\bf B521} (2001) 29--32,
  [\href{http://arxiv.org/abs/hep-ph/0102289}{{\tt hep-ph/0102289}}].

\bibitem{Buchler:2001nm}
M.~B{\"u}chler, G.~Colangelo, J.~Kambor, and F.~Orellana, {\it {Dispersion
  relations and soft pion theorems for $K\to\pi\pi$}},  {\em Phys. Lett.} {\bf
  B521} (2001) 22--28, [\href{http://arxiv.org/abs/hep-ph/0102287}{{\tt
  hep-ph/0102287}}].

\bibitem{Pallante:2001he}
E.~Pallante, A.~Pich, and I.~Scimemi, {\it {The Standard model prediction for
  $\varepsilon'/\varepsilon$}},  {\em Nucl. Phys.} {\bf B617} (2001) 441--474,
  [\href{http://arxiv.org/abs/hep-ph/0105011}{{\tt hep-ph/0105011}}].

\bibitem{Buras:2000kx}
A.~J. Buras et~al., {\it {Final state interactions and $\epe$: A critical
  look}},  {\em Phys.~Lett.} {\bf B480} (2000) 80--86,
  [\href{http://arxiv.org/abs/hep-ph/0002116}{{\tt hep-ph/0002116}}].

\bibitem{Cirigliano:2003nn}
V.~Cirigliano, A.~Pich, G.~Ecker, and H.~Neufeld, {\it {Isospin violation in
  $\epsilon^\prime$}},  {\em Phys.~Rev.~Lett.} {\bf 91} (2003) 162001,
  [\href{http://arxiv.org/abs/hep-ph/0307030}{{\tt hep-ph/0307030}}].

\bibitem{Cirigliano:2003gt}
V.~Cirigliano, G.~Ecker, H.~Neufeld, and A.~Pich, {\it {Isospin breaking in
  $K\to\pi\pi$ decays}},  {\em Eur. Phys. J.} {\bf C33} (2004) 369--396,
  [\href{http://arxiv.org/abs/hep-ph/0310351}{{\tt hep-ph/0310351}}].

\bibitem{Bijnens:2004ai}
J.~Bijnens and F.~Borg, {\it {Isospin breaking in $K\to 3 \pi$ decays III:
  Bremsstrahlung and fit to experiment}},  {\em Eur. Phys. J.} {\bf C40} (2005)
  383--394, [\href{http://arxiv.org/abs/hep-ph/0501163}{{\tt hep-ph/0501163}}].

\bibitem{Bertolini:1998vd}
S.~Bertolini, M.~Fabbrichesi, and J.~O. Eeg, {\it {Theory of the CP violating
  parameter $\epsilon'/\epsilon$}},  {\em Rev.~Mod.~Phys.} {\bf 72} (2000)
  65--93, [\href{http://arxiv.org/abs/hep-ph/9802405}{{\tt hep-ph/9802405}}].

\bibitem{Cirigliano:2011ny}
V.~Cirigliano, G.~Ecker, H.~Neufeld, A.~Pich, and J.~Portoles, {\it {Kaon
  Decays in the Standard Model}},  {\em Rev.~Mod.~Phys.} {\bf 84} (2012) 399,
  [\href{http://arxiv.org/abs/1107.6001}{{\tt arXiv:1107.6001}}].

\bibitem{Bertolini:2012pu}
S.~Bertolini, J.~O. Eeg, A.~Maiezza, and F.~Nesti, {\it {New physics in
  $\epsilon'$ from gluomagnetic contributions and limits on Left-Right
  symmetry}},  {\em Phys.~Rev.} {\bf D86} (2012) 095013,
  [\href{http://arxiv.org/abs/1206.0668}{{\tt arXiv:1206.0668}}].

\bibitem{Bai:2015nea}
{\bf RBC, UKQCD} Collaboration, Z.~Bai et~al., {\it {Standard Model Prediction
  for Direct CP Violation in $K\to\pi\pi$ Decay}},  {\em Phys. Rev. Lett.} {\bf
  115} (2015), no.~21 212001, [\href{http://arxiv.org/abs/1505.07863}{{\tt
  arXiv:1505.07863}}].

\bibitem{Blum:2015ywa}
T.~Blum et~al., {\it {$K \rightarrow \pi\pi$ $\Delta I=3/2$ decay amplitude in
  the continuum limit}},  {\em Phys.~Rev.} {\bf D91} (2015), no.~7 074502,
  [\href{http://arxiv.org/abs/1502.00263}{{\tt arXiv:1502.00263}}].

\bibitem{Kitahara:2016nld}
T.~Kitahara, U.~Nierste, and P.~Tremper, {\it {Singularity-free next-to-leading
  order $\Delta$S = 1 renormalization group evolution and
  $\epsilon_K'/\epsilon_K$ in the Standard Model and beyond}},  {\em JHEP} {\bf
  12} (2016) 078, [\href{http://arxiv.org/abs/1607.06727}{{\tt
  arXiv:1607.06727}}].

\bibitem{Buras:2015xba}
A.~J. Buras and J.-M. G\'erard, {\it {Upper Bounds on
  $\varepsilon'/\varepsilon$ Parameters $B_6^{(1/2)}$ and $B_8^{(3/2)}$ from
  Large N QCD and other News}},  {\em JHEP} {\bf 12} (2015) 008,
  [\href{http://arxiv.org/abs/1507.06326}{{\tt arXiv:1507.06326}}].

\bibitem{Buras:2014maa}
A.~J. Buras, J.-M. G{\'e}rard, and W.~A. Bardeen, {\it {Large $N$ Approach to
  Kaon Decays and Mixing 28 Years Later: $\Delta I = 1/2$ Rule, $\hat B_K$ and
  $\Delta M_K$}},  {\em Eur.~Phys.~J.} {\bf C74} (2014), no.~5 2871,
  [\href{http://arxiv.org/abs/1401.1385}{{\tt arXiv:1401.1385}}].

\bibitem{Colangelo:2001df}
G.~Colangelo, J.~Gasser, and H.~Leutwyler, {\it {$\pi \pi$ scattering}},  {\em
  Nucl. Phys.} {\bf B603} (2001) 125--179,
  [\href{http://arxiv.org/abs/hep-ph/0103088}{{\tt hep-ph/0103088}}].

\bibitem{Gisbert:2017vvj}
H.~Gisbert and A.~Pich, {\it {Direct CP violation in $K^0\to\pi\pi$: Standard
  Model Status}},  {\em Rept. Prog. Phys.} {\bf 81} (2018), no.~7 076201,
  [\href{http://arxiv.org/abs/1712.06147}{{\tt arXiv:1712.06147}}].

\bibitem{Gisbert:2018tuf}
H.~Gisbert and A.~Pich, {\it {Updated Standard Model Prediction for
  $\varepsilon'/\varepsilon$}},  {\em Nucl. Part. Phys. Proc.} {\bf 300-302}
  (2018) 137--144, [\href{http://arxiv.org/abs/1810.04904}{{\tt
  arXiv:1810.04904}}].

\bibitem{Gisbert:2018niu}
H.~Gisbert, {\it {Current status of $\varepsilon'/\varepsilon$ in the Standard
  Model}},  {\em PoS} {\bf Confinement2018} (2018) 178,
  [\href{http://arxiv.org/abs/1811.12206}{{\tt arXiv:1811.12206}}].

\bibitem{Buras:2016fys}
A.~J. Buras and J.-M. G{\'e}rard, {\it {Final state interactions in
  $K\rightarrow \pi \pi $ decays: $\Delta I=1/2$ rule vs. $\varepsilon
  '/\varepsilon $}},  {\em Eur. Phys. J.} {\bf C77} (2017), no.~1 10,
  [\href{http://arxiv.org/abs/1603.05686}{{\tt arXiv:1603.05686}}].

\bibitem{Buras:2018ozh}
A.~J. Buras, {\it {$\epsilon^\prime/\epsilon$-2018: A Christmas Story}},
  \href{http://arxiv.org/abs/1812.06102}{{\tt arXiv:1812.06102}}.

\bibitem{Cirigliano:2019cpi}
V.~Cirigliano, H.~Gisbert, A.~Pich, and A.~Rodríguez-Sánchez, {\it
  {Isospin-violating contributions to $\epsilon'/\epsilon$}},  {\em JHEP} {\bf
  02} (2020) 032, [\href{http://arxiv.org/abs/1911.01359}{{\tt
  arXiv:1911.01359}}].

\bibitem{Cirigliano:2019ani}
V.~Cirigliano, H.~Gisbert, A.~Pich, and A.~Rodr\'\i{}guez-S\'anchez, {\it
  {Theoretical status of $\varepsilon'/\varepsilon$}},  {\em J. Phys. Conf.
  Ser.} {\bf 1526} (2020) 012011, [\href{http://arxiv.org/abs/1912.04736}{{\tt
  arXiv:1912.04736}}].

\bibitem{Gisbert:2020wkb}
H.~Gisbert, {\it {Nonleptonic $K\to 2\,\pi$ decay dynamics}},  12, 2020.
\newblock \href{http://arxiv.org/abs/2012.09483}{{\tt arXiv:2012.09483}}.

\bibitem{Bobeth:2017ecx}
C.~Bobeth and A.~J. Buras, {\it {Leptoquarks meet $\varepsilon'/\varepsilon$
  and rare Kaon processes}},  {\em JHEP} {\bf 02} (2018) 101,
  [\href{http://arxiv.org/abs/1712.01295}{{\tt arXiv:1712.01295}}].

\bibitem{Aebischer:2018csl}
J.~Aebischer, C.~Bobeth, A.~J. Buras, and D.~M. Straub, {\it {Anatomy of
  $\varepsilon'/\varepsilon$ beyond the Standard Model}},  {\em Eur. Phys. J.}
  {\bf C79} (2019), no.~3 219, [\href{http://arxiv.org/abs/1808.00466}{{\tt
  arXiv:1808.00466}}].

\bibitem{Aebischer:2018quc}
J.~Aebischer, C.~Bobeth, A.~J. Buras, J.-M. G{\'e}rard, and D.~M. Straub, {\it
  {Master formula for $\varepsilon'/\varepsilon$ beyond the Standard Model}},
  {\em Phys. Lett.} {\bf B792} (2019) 465--469,
  [\href{http://arxiv.org/abs/1807.02520}{{\tt arXiv:1807.02520}}].

\bibitem{Aebischer:2018rrz}
J.~Aebischer, A.~J. Buras, and J.-M. G{\'e}rard, {\it {BSM Hadronic Matrix
  Elements for $\epe$ and $K\to\pi\pi$ Decays in the Dual QCD Approach}},  {\em
  JHEP} {\bf 02} (2019) 021, [\href{http://arxiv.org/abs/1807.01709}{{\tt
  arXiv:1807.01709}}].

\bibitem{Constantinou:2017sgv}
{\bf ETM} Collaboration, M.~Constantinou, M.~Costa, R.~Frezzotti, V.~Lubicz,
  G.~Martinelli, D.~Meloni, H.~Panagopoulos, and S.~Simula, {\it {$K \to \pi$
  matrix elements of the chromomagnetic operator on the lattice}},  {\em Phys.
  Rev.} {\bf D97} (2018), no.~7 074501,
  [\href{http://arxiv.org/abs/1712.09824}{{\tt arXiv:1712.09824}}].

\bibitem{Buras:2018evv}
A.~J. Buras and J.-M. G{\'e}rard, {\it {$K\to\pi\pi$ and $K-\pi$ Matrix
  Elements of the Chromomagnetic Operators from Dual QCD}},  {\em JHEP} {\bf
  07} (2018) 126, [\href{http://arxiv.org/abs/1803.08052}{{\tt
  arXiv:1803.08052}}].

\bibitem{Buras:2020pjp}
A.~J. Buras and J.-M. G\'erard, {\it {Isospin-breaking in $\varepsilon
  '/\varepsilon $: impact of $\eta _0$ at the dawn of the 2020s}},  {\em Eur.
  Phys. J. C} {\bf 80} (2020), no.~8 701,
  [\href{http://arxiv.org/abs/2005.08976}{{\tt arXiv:2005.08976}}].

\bibitem{Buras:2019vik}
A.~J. Buras, {\it {The Optimal Strategy for $\varepsilon'/\varepsilon$ in the
  SM: 2019}},  {\em J. Phys. Conf. Ser.} {\bf 1526} (2020) 012019,
  [\href{http://arxiv.org/abs/1912.12306}{{\tt arXiv:1912.12306}}].

\bibitem{Ishizuka:2018qbn}
N.~Ishizuka, K.-I. Ishikawa, A.~Ukawa, and T.~Yoshi\'e, {\it {Calculation of $K
  \to \pi\pi$ decay amplitudes with improved Wilson fermion action in non-zero
  momentum frame in lattice QCD}},  {\em Phys. Rev. D} {\bf 98} (2018), no.~11
  114512, [\href{http://arxiv.org/abs/1809.03893}{{\tt arXiv:1809.03893}}].

\bibitem{GellMann:1955jx}
M.~Gell-Mann and A.~Pais, {\it {Behavior of neutral particles under charge
  conjugation}},  {\em Phys.~Rev.} {\bf 97} (1955) 1387--1389.

\bibitem{GellMann:1957wh}
M.~Gell-Mann and A.~Rosenfeld, {\it {Hyperons and heavy mesons (systematics and
  decay)}},  {\em Ann.Rev.Nucl.Part.Sci.} {\bf 7} (1957) 407--478.

\bibitem{Donini:2020qfu}
A.~Donini, P.~Hern\'andez, C.~Pena, and F.~Romero-L\'opez, {\it {Dissecting the
  $\Delta I= 1/2$ rule at large $N_c$}},  {\em Eur. Phys. J. C} {\bf 80}
  (2020), no.~7 638, [\href{http://arxiv.org/abs/2003.10293}{{\tt
  arXiv:2003.10293}}].

\bibitem{Hernandez:2020tbc}
P.~Hern\'andez and F.~Romero-L\'opez, {\it {The Large $N_c$ limit of QCD on the
  lattice}},  12, 2020.
\newblock \href{http://arxiv.org/abs/2012.03331}{{\tt arXiv:2012.03331}}.

\bibitem{Chivukula:1986du}
R.~S. Chivukula, J.~Flynn, and H.~Georgi, {\it {Polychromatic Penguins Don't
  Fly}},  {\em Phys.Lett.} {\bf B171} (1986) 453--458.

\bibitem{Pich:1990mw}
A.~Pich and E.~de~Rafael, {\it {Four quark operators and nonleptonic weak
  transitions}},  {\em Nucl. Phys. B} {\bf 358} (1991) 311--382.

\bibitem{Jamin:1994sv}
M.~Jamin and A.~Pich, {\it {QCD corrections to inclusive Delta S = 1,2
  transitions at the next-to-leading order}},  {\em Nucl. Phys. B} {\bf 425}
  (1994) 15--38, [\href{http://arxiv.org/abs/hep-ph/9402363}{{\tt
  hep-ph/9402363}}].

\bibitem{Pich:1995qp}
A.~Pich and E.~de~Rafael, {\it {Weak K amplitudes in the chiral and $1/N$
  expansions}},  {\em Phys.Lett.} {\bf B374} (1996) 186--192,
  [\href{http://arxiv.org/abs/hep-ph/9511465}{{\tt hep-ph/9511465}}].

\bibitem{Bertolini:1997ir}
S.~Bertolini, J.~Eeg, M.~Fabbrichesi, and E.~Lashin, {\it {The $\Delta I = 1/2$
  rule and B(K) at $\ord(p^4)$ in the chiral expansion}},  {\em Nucl. Phys. B}
  {\bf 514} (1998) 63--92, [\href{http://arxiv.org/abs/hep-ph/9705244}{{\tt
  hep-ph/9705244}}].

\bibitem{Bijnens:1998ee}
J.~Bijnens and J.~Prades, {\it {The $\Delta I = 1/2$ rule in the chiral
  limit}},  {\em JHEP} {\bf 01} (1999) 023,
  [\href{http://arxiv.org/abs/hep-ph/9811472}{{\tt hep-ph/9811472}}].

\bibitem{Hambye:1998sma}
T.~Hambye, G.~Kohler, E.~Paschos, P.~Soldan, and W.~A. Bardeen, {\it {$1 / N$
  corrections to the hadronic matrix elements of $Q_6$ and $Q_8$ in $K\to\pi
  \pi$ decays}},  {\em Phys.~Rev.} {\bf D58} (1998) 014017,
  [\href{http://arxiv.org/abs/hep-ph/9802300}{{\tt hep-ph/9802300}}].

\bibitem{Crewther:2013vea}
R.~Crewther and L.~C. Tunstall, {\it {$\Delta I=1/2$ rule for kaon decays
  derived from QCD infrared fixed point}},  {\em Phys. Rev. D} {\bf 91} (2015),
  no.~3 034016, [\href{http://arxiv.org/abs/1312.3319}{{\tt arXiv:1312.3319}}].

\bibitem{Tanabashi:2018oca}
{\bf Particle Data Group} Collaboration, M.~Tanabashi et~al., {\it {Review of
  Particle Physics}},  {\em Phys. Rev.} {\bf D98} (2018), no.~3 030001.

\bibitem{Buras:2018lgu}
A.~J. Buras and J.-M. G{\'e}rard, {\it {Dual QCD Insight into BSM Hadronic
  Matrix Elements for $K^0-\bar K^0$ Mixing from Lattice QCD}},  {\em Acta
  Phys. Polon.} {\bf B50} (2019) 121,
  [\href{http://arxiv.org/abs/1804.02401}{{\tt arXiv:1804.02401}}].

\bibitem{Carrasco:2015pra}
{\bf ETM} Collaboration, N.~Carrasco, P.~Dimopoulos, R.~Frezzotti, V.~Lubicz,
  G.~C. Rossi, S.~Simula, and C.~Tarantino, {\it {$\Delta S=2$ and $\Delta C=2$
  bag parameters in the standard model and beyond from N$_f$=2+1+1 twisted-mass
  lattice QCD}},  {\em Phys. Rev.} {\bf D92} (2015), no.~3 034516,
  [\href{http://arxiv.org/abs/1505.06639}{{\tt arXiv:1505.06639}}].

\bibitem{Jang:2015sla}
{\bf SWME} Collaboration, B.~J. Choi et~al., {\it {Kaon BSM B-parameters using
  improved staggered fermions from $N_f=2+1$ unquenched QCD}},  {\em Phys.
  Rev.} {\bf D93} (2016), no.~1 014511,
  [\href{http://arxiv.org/abs/1509.00592}{{\tt arXiv:1509.00592}}].

\bibitem{Garron:2016mva}
{\bf RBC/UKQCD} Collaboration, N.~Garron, R.~J. Hudspith, and A.~T. Lytle, {\it
  {Neutral Kaon Mixing Beyond the Standard Model with $n_f=2+1$ Chiral Fermions
  Part 1: Bare Matrix Elements and Physical Results}},  {\em JHEP} {\bf 11}
  (2016) 001, [\href{http://arxiv.org/abs/1609.03334}{{\tt arXiv:1609.03334}}].

\bibitem{Boyle:2017skn}
{\bf RBC, UKQCD} Collaboration, P.~A. Boyle, N.~Garron, R.~J. Hudspith,
  C.~Lehner, and A.~T. Lytle, {\it {Neutral kaon mixing beyond the Standard
  Model with n$_{f}$ = 2 + 1 chiral fermions. Part 2: non perturbative
  renormalisation of the $\Delta F=2$ four-quark operators}},  {\em JHEP} {\bf
  10} (2017) 054, [\href{http://arxiv.org/abs/1708.03552}{{\tt
  arXiv:1708.03552}}].

\bibitem{Boyle:2017ssm}
P.~Boyle, N.~Garron, J.~Kettle, A.~Khamseh, and J.~T. Tsang, {\it {BSM Kaon
  Mixing at the Physical Point}},  {\em EPJ Web Conf.} {\bf 175} (2018) 13010,
  [\href{http://arxiv.org/abs/1710.09176}{{\tt arXiv:1710.09176}}].

\bibitem{Brown:1990tt}
G.~Brown, J.~Durso, M.~Johnson, and J.~Speth, {\it {Final state interactions in
  K meson decays}},  {\em Phys. Lett. B} {\bf 238} (1990) 20--24.

\bibitem{'tHooft:1973jz}
G.~'t~Hooft, {\it {A Planar Diagram Theory for Strong Interactions}},  {\em
  Nucl.~Phys.} {\bf B72} (1974) 461.

\bibitem{'tHooft:1974hx}
G.~'t~Hooft, {\it {A Two-Dimensional Model for Mesons}},  {\em Nucl.~Phys.}
  {\bf B75} (1974) 461.

\bibitem{Witten:1979kh}
E.~Witten, {\it {Baryons in the $1/N$ Expansion}},  {\em Nucl.~Phys.} {\bf
  B160} (1979) 57.

\bibitem{Treiman:1986ep}
S.~B. Treiman, E.~Witten, R.~Jackiw, and B.~Zumino, {\em {Current Algebra and
  Anomalies}}.
\newblock 1986.

\bibitem{Buras:2014sba}
A.~J. Buras, F.~De~Fazio, and J.~Girrbach, {\it {$\Delta I=1/2$ rule,
  $\varepsilon '/\varepsilon $ and $K\rightarrow \pi \nu \bar{\nu }$ in $Z'
  (Z)$ and $G' $ models with FCNC quark couplings}},  {\em Eur.~Phys.~J.} {\bf
  C74} (2014) 2950, [\href{http://arxiv.org/abs/1404.3824}{{\tt
  arXiv:1404.3824}}].

\bibitem{Buras:2015jaq}
A.~J. Buras, {\it {New physics patterns in $\varepsilon^\prime/\varepsilon$ and
  $\varepsilon_K$ with implications for rare kaon decays and $\Delta M_K$}},
  {\em JHEP} {\bf 04} (2016) 071, [\href{http://arxiv.org/abs/1601.00005}{{\tt
  arXiv:1601.00005}}].

\bibitem{Aebischer:2020mkv}
J.~Aebischer, A.~J. Buras, and J.~Kumar, {\it {Another SMEFT Story: $Z^\prime$
  Facing New Results on $\varepsilon'/\varepsilon$, $\Delta M_K$ and
  $K\to\pi\nu\bar\nu$}},  {\em JHEP} {\bf 12} (2020) 097,
  [\href{http://arxiv.org/abs/2006.01138}{{\tt arXiv:2006.01138}}].

\bibitem{Blanke:2015wba}
M.~Blanke, A.~J. Buras, and S.~Recksiegel, {\it {Quark flavour observables in
  the Littlest Higgs model with T-parity after LHC Run 1}},  {\em Eur. Phys.
  J.} {\bf C76} (2016), no.~4 182, [\href{http://arxiv.org/abs/1507.06316}{{\tt
  arXiv:1507.06316}}].

\bibitem{Bobeth:2017xry}
C.~Bobeth, A.~J. Buras, A.~Celis, and M.~Jung, {\it {Yukawa enhancement of
  $Z$-mediated new physics in $\Delta S = 2$ and $\Delta B = 2$ processes}},
  {\em JHEP} {\bf 07} (2017) 124, [\href{http://arxiv.org/abs/1703.04753}{{\tt
  arXiv:1703.04753}}].

\bibitem{Endo:2016tnu}
M.~Endo, T.~Kitahara, S.~Mishima, and K.~Yamamoto, {\it {Revisiting Kaon
  Physics in General $Z$ Scenario}},  {\em Phys. Lett.} {\bf B771} (2017)
  37--44, [\href{http://arxiv.org/abs/1612.08839}{{\tt arXiv:1612.08839}}].

\bibitem{Buras:2015yca}
A.~J. Buras, D.~Buttazzo, and R.~Knegjens, {\it {$K\to\pi\nu\bar\nu$ and
  $\epsilon'/\epsilon$ in Simplified New Physics Models}},  {\em JHEP} {\bf 11}
  (2015) 166, [\href{http://arxiv.org/abs/1507.08672}{{\tt arXiv:1507.08672}}].

\bibitem{Buras:2015kwd}
A.~J. Buras and F.~De~Fazio, {\it {$\varepsilon'/\varepsilon$ in 331 Models}},
  {\em JHEP} {\bf 03} (2016) 010, [\href{http://arxiv.org/abs/1512.02869}{{\tt
  arXiv:1512.02869}}].

\bibitem{Buras:2016dxz}
A.~J. Buras and F.~De~Fazio, {\it {331 Models Facing the Tensions in $\Delta
  F=2$ Processes with the Impact on $\varepsilon^\prime/\varepsilon$,
  $B_s\to\mu^+\mu^-$ and $B\to K^*\mu^+\mu^-$}},  {\em JHEP} {\bf 08} (2016)
  115, [\href{http://arxiv.org/abs/1604.02344}{{\tt arXiv:1604.02344}}].

\bibitem{Bobeth:2016llm}
C.~Bobeth, A.~J. Buras, A.~Celis, and M.~Jung, {\it {Patterns of Flavour
  Violation in Models with Vector-Like Quarks}},  {\em JHEP} {\bf 04} (2017)
  079, [\href{http://arxiv.org/abs/1609.04783}{{\tt arXiv:1609.04783}}].

\bibitem{Tanimoto:2016yfy}
M.~Tanimoto and K.~Yamamoto, {\it {Probing SUSY with 10 TeV stop mass in rare
  decays and CP violation of kaon}},  {\em PTEP} {\bf 2016} (2016), no.~12
  123B02, [\href{http://arxiv.org/abs/1603.07960}{{\tt arXiv:1603.07960}}].

\bibitem{Kitahara:2016otd}
T.~Kitahara, U.~Nierste, and P.~Tremper, {\it {Supersymmetric Explanation of CP
  Violation in $K\to \pi\pi$ Decays}},  {\em Phys. Rev. Lett.} {\bf 117}
  (2016), no.~9 091802, [\href{http://arxiv.org/abs/1604.07400}{{\tt
  arXiv:1604.07400}}].

\bibitem{Endo:2016aws}
M.~Endo, S.~Mishima, D.~Ueda, and K.~Yamamoto, {\it {Chargino contributions in
  light of recent $\epsilon'/\epsilon$}},  {\em Phys. Lett.} {\bf B762} (2016)
  493--497, [\href{http://arxiv.org/abs/1608.01444}{{\tt arXiv:1608.01444}}].

\bibitem{Crivellin:2017gks}
A.~Crivellin, G.~D'Ambrosio, T.~Kitahara, and U.~Nierste, {\it {$K\to \pi
  \nu\overline{\nu}$ in the MSSM in light of the
  $\epsilon^{\prime}_K/\epsilon_K$ anomaly}},  {\em Phys. Rev.} {\bf D96}
  (2017), no.~1 015023, [\href{http://arxiv.org/abs/1703.05786}{{\tt
  arXiv:1703.05786}}].

\bibitem{Endo:2017ums}
M.~Endo, T.~Goto, T.~Kitahara, S.~Mishima, D.~Ueda, and K.~Yamamoto, {\it
  {Gluino-mediated electroweak penguin with flavor-violating trilinear
  couplings}},  {\em JHEP} {\bf 04} (2018) 019,
  [\href{http://arxiv.org/abs/1712.04959}{{\tt arXiv:1712.04959}}].

\bibitem{Chen:2018ytc}
C.-H. Chen and T.~Nomura, {\it {$Re(\epsilon'_K/\epsilon_K$) and $K \to \pi \nu
  \bar\nu$ in a two-Higgs doublet model}},  {\em JHEP} {\bf 08} (2018) 145,
  [\href{http://arxiv.org/abs/1804.06017}{{\tt arXiv:1804.06017}}].

\bibitem{Chen:2018vog}
C.-H. Chen and T.~Nomura, {\it {$\epsilon'/\epsilon$ from charged-Higgs-induced
  gluonic dipole operators}},  {\em Phys. Lett.} {\bf B787} (2018) 182--187,
  [\href{http://arxiv.org/abs/1805.07522}{{\tt arXiv:1805.07522}}].

\bibitem{Cirigliano:2016yhc}
V.~Cirigliano, W.~Dekens, J.~de~Vries, and E.~Mereghetti, {\it {An
  $\varepsilon'$ improvement from right-handed currents}},  {\em Phys. Lett.}
  {\bf B767} (2017) 1--9, [\href{http://arxiv.org/abs/1612.03914}{{\tt
  arXiv:1612.03914}}].

\bibitem{Alioli:2017ces}
S.~Alioli, V.~Cirigliano, W.~Dekens, J.~de~Vries, and E.~Mereghetti, {\it
  {Right-handed charged currents in the era of the Large Hadron Collider}},
  {\em JHEP} {\bf 05} (2017) 086, [\href{http://arxiv.org/abs/1703.04751}{{\tt
  arXiv:1703.04751}}].

\bibitem{Haba:2018byj}
N.~Haba, H.~Umeeda, and T.~Yamada, {\it {$\epsilon'/\epsilon$ Anomaly and
  Neutron EDM in $SU(2)_L\times SU(2)_R\times U(1)_{B-L}$ model with Charge
  Symmetry}},  {\em JHEP} {\bf 05} (2018) 052,
  [\href{http://arxiv.org/abs/1802.09903}{{\tt arXiv:1802.09903}}].

\bibitem{Haba:2018rzf}
N.~Haba, H.~Umeeda, and T.~Yamada, {\it {Direct CP Violation in Cabibbo-Favored
  Charmed Meson Decays and $\epsilon'/\epsilon$ in $SU(2)_L\times SU(2)_R\times
  U(1)_{B-L}$ Model}},  {\em JHEP} {\bf 10} (2018) 006,
  [\href{http://arxiv.org/abs/1806.03424}{{\tt arXiv:1806.03424}}].

\bibitem{Matsuzaki:2018jui}
S.~Matsuzaki, K.~Nishiwaki, and K.~Yamamoto, {\it {Simultaneous interpretation
  of $K$ and $B$ anomalies in terms of chiral-flavorful vectors}},  {\em JHEP}
  {\bf 11} (2018) 164, [\href{http://arxiv.org/abs/1806.02312}{{\tt
  arXiv:1806.02312}}].

\bibitem{Chen:2018dfc}
C.-H. Chen and T.~Nomura, {\it {$\epsilon_K$ and $\epsilon'/\epsilon$ in a
  diquark model}},  {\em JHEP} {\bf 03} (2019) 009,
  [\href{http://arxiv.org/abs/1808.04097}{{\tt arXiv:1808.04097}}].

\bibitem{Chen:2018stt}
C.-H. Chen and T.~Nomura, {\it {Left-handed color-sextet diquark in the Kaon
  system}},  {\em Phys. Rev.} {\bf D99} (2019), no.~11 115006,
  [\href{http://arxiv.org/abs/1811.02315}{{\tt arXiv:1811.02315}}].

\bibitem{Marzo:2019ldg}
C.~Marzo, L.~Marzola, and M.~Raidal, {\it {Common explanation to the
  $R_{K^{(*)}}$, $R_{D^{(*)}}$ and $\epsilon^\prime/\epsilon$ anomalies in a
  3HDM+$\nu_R$ and connections to neutrino physics}},  {\em Phys. Rev.} {\bf
  D100} (2019), no.~5 055031, [\href{http://arxiv.org/abs/1901.08290}{{\tt
  arXiv:1901.08290}}].

\bibitem{Matsuzaki:2019clv}
S.~Matsuzaki, K.~Nishiwaki, and K.~Yamamoto, {\it {Simultaneous explanation of
  $K$ and $B$ anomalies in vectorlike compositeness}},  in {\em {18th Hellenic
  School and Workshops on Elementary Particle Physics and Gravity (CORFU2018)
  Corfu, Corfu, Greece, August 31-September 28, 2018}}, 2019.
\newblock \href{http://arxiv.org/abs/1903.10823}{{\tt arXiv:1903.10823}}.

\bibitem{Crivellin:2019isj}
A.~Crivellin, C.~Gross, S.~Pokorski, and L.~Vernazza, {\it {Correlating
  $\epsilon^\prime/\epsilon$ to hadronic $B$ decays via $U(2)^3$ flavour
  symmetry}},  {\em Phys. Rev. D} {\bf 101} (2020), no.~1 015022,
  [\href{http://arxiv.org/abs/1909.02101}{{\tt arXiv:1909.02101}}].

\end{thebibliography}\endgroup
\end{document}